\documentclass[preprint,eqsecnum,aps,tightenlines,showpacs,
floats]{revtex4}
\usepackage{graphics}
\usepackage{graphicx}
\usepackage{epsf} 
\usepackage{epsfig}

\begin{document}


\def\Xint#1{\mathchoice
   {\XXint\displaystyle\textstyle{#1}}%
   {\XXint\textstyle\scriptstyle{#1}}%
   {\XXint\scriptstyle\scriptscriptstyle{#1}}%
   {\XXint\scriptscriptstyle\scriptscriptstyle{#1}}%
   \!\int}
\def\XXint#1#2#3{{\setbox0=\hbox{$#1{#2#3}{\int}$}
     \vcenter{\hbox{$#2#3$}}\kern-.5\wd0}}
\def\ddashint{\Xint=}
\def\dashint{\Xint-}


\preprint{
\parbox{1.5in}{ \leftline{JLAB-THY-03-195}
                \leftline{WM-03-109}
                 }}

\title
{\bf \ \\ Quark-Hadron Duality and Scaling in Reduced
QCD}
\author{
Zolt$\acute{{\rm a}}$n Batiz$^{1}$ and Franz Gross$^{2,3}$} 
\affiliation{
$^1$Centro de F\'isica das Interac\c{c}\~oes Fundamentais
(CFIF), Instituto Superior T\'ecnico,
 P-1049-001 Lisboa, Portugal \\
$^2$Department of Physics, College of William and Mary,
Williamsburg, Virginia 23185\\
$^3$Thomas Jefferson National Accelerator Facility, Newport
News,  Virginia 23606}
\date{\today }

\begin{abstract} 
We introduce a generalization of $1+1$ dimensional large $N_c$
QCD, which we refer to as ``reduced'' QCD, or rQCD. In this
model gluons and quark momenta live in $1+1$ dimensions only, but the quark spin and all other
particles (leptons, and the photon) live in the full
$1+3$ dimensions.  The bound states of quarks and
antiquarks are identical to those originally described by 't
Hooft (except that there are new transversely
polarized states previously excluded), so the model is exactly
soluable. However, significant differences arise when the model
is applied to electromagnetic interactions.  After reviewing
the strongly interacting sector of the theory, we discuss deep
inelastic scattering (DIS) in this model, and show that the new
states with transverse polarization give the Callan-Gross
relation and remove the pathological features of
the original 1+1 dimensional description.  We conclude that rQCD 
gives a satisfactory description of the phenomenology and
provides a deep understanding of both duality and
DIS.
\end{abstract}


 

\phantom{0}
\vspace{5.5in}

\section{Introduction}
\label{sec:introduction}

In 1974, 't Hooft described the behavior of the large
$N_c$ limit of QCD in $1+1$ dimensions \cite{thooft}.  He
showed that the model confined quarks and was exactly
soluable.   His work was followed quickly by several papers  
\cite{Callan:ps,EINHORN,BG} discussing various aspects of the
model.  The work by Einhorn \cite{EINHORN} discussed the 't
Hooft model predictions for deep inelastic scattering (DIS). 
Einhorn found that the DIS cross section was proportional to the
square of the undressed quark masses, and hence would
approach zero as the quark masses approach zero.  This
pathlogical result made it difficult to apply the 't Hooft
model to the phenomenology of DIS; clearly someting major was
missing.  Of course, limiting the physics to
$1+1$ dimensions also excludes transverse spin degrees of
freedom, so $W_1$ was zero and the Callan-Gross relation
\cite{Callan:uq} did not hold.  

Since then QCD in $1+1$ diminsions has been studied from a
variety of points of view
\cite{Burkardt:2000uu,Burkardt:2000ez,Schon:2000qy,
Burkardt:2002yf, Kalashnikova:2001df} and applied recently to
the study of duality in heavy meson decays
\cite{gl,l,lu}.  However, to the best of our knowledge, no one
has followed up on Einhorn's study \cite{EINHORN} of DIS. 

\vspace{-9.0in}
\maketitle 

A few years ago we found that a very simple toy model of scalar
and spinor fields in $1+1$ dimension gave nice
phenomenological results for DIS \cite{ref1}.  By extending
the spinor degrees of freedom to 1+3 dimensions we were able to
recover the Callan-Gross relation.  We decided to extend our
approach to the 't Hooft model, and learned of the
pathology in its description of DIS.  It turns out that
this pathology can be removed by extending the spinor
degrees of freedom to 1+3 dimensions, just as we did before,
and this paper grew out of that study.    
 
This paper is divided into five sections, with seven appendices that include many of the details.  Following this brief
introduction, Sec.\ II reviews the 't Hooft model results for
the strong sector.  Our discussion focuses on (i) a method of 
treatment of the confining interaction that removes all
singularities from the theory and gives finite dressed quark
masses, and (ii) the consequence of including transverse degrees
of freedom for the spin of the quarks.  We calculate the dressed
quark mass, and the properties and spectrum of $q\bar q$ bound
states, and also study the consequences of the completness of
the bound states.  We show how to construct an off-shell
$q\bar q$ scattering matrix, and show that the on-shell
scattering matrix must be zero, as required for confined
particles.  Sections III and IV study 
electromagnetic interactions and deep inelastic scatterng.  Here
the presence of transverse spin degrees of freedom completly
alters the Einhorn discussion, giving us phenomonologically
useful results. We conclude with a brief discussion.  

In a subsequent paper \cite{BGII} we plan to present a numerical study of the approach to scaling in the DIS limit.

\section{the `t Hooft Model}

\subsection{The Lagrangian and confining interaction}

We propose that the action have the following form
\begin{equation}
{\cal S}={\cal S}_{QCD} + {\cal S}_{QED}
\label{A1}
\end{equation}
where the QCD part of the action (gluons and quark momenta or coordinates) live in $1+1$ dimensions only, but the quark spin and the QED part (leptons and the photon) live in the full
$1+3$ dimensions 
\begin{eqnarray}
{\cal S}_{QCD} &=&\int dt\,dz \; {\cal L}_{QCD}(t,z) \nonumber\\
{\cal S}_{QED} &=&\int dt\,d{\bf r} \; {\cal L}_{QED}(t,{\bf
r})\, . 
\label{A2}
\end{eqnarray}
In his original paper, 't Hooft \cite{thooft}
discussed the strong interactions only, and in this section we
review the 1+1 dimensional model of QCD that he presented. The
QED part of the action (\ref{A2}), and reduced QCD (rQCD), will
be discussed in Sec.~\ref{RQCD}.  

The QCD Lagrangian density is
\begin{equation}
{\cal L}_{QCD}(t,z)=-\frac{1}{4}Tr \left[ F^{\mu \nu}F_{\mu \nu}
\right]+ \sum_i
\bar{q}_i\left(i D_{\mu}
\gamma^{\mu}-m_{0i}
\right)q_i\, ,
\label{1eq1}
\end{equation}
where $q_i=q_i(t,z)$ is the quark field with flavor $i$ and bare
mass $m_{0i}$.  The gluon field quantities are 
\begin{eqnarray}
A^{\mu}&=&\frac{1}{2}\,A^{\mu}_a \lambda_a \\ \nonumber
F_{\mu \nu}&=&\partial_{\mu} A_{\nu}-\partial_{\nu} A_{\mu}
+ig_0 \left[ A_{\mu},A_{\nu} \right] \\ \nonumber
D_{\mu}&=&\partial_{\mu}+ig_0 A_{\mu},
\label{1eq2}
\end{eqnarray} 
where $F_{\mu\nu}$ is the gluon field tensor and
$A^{\mu}_{a}=A^{\mu}_{a}(t,z)$ are the gluon fields with the
Lorentz index $\mu$ and the color index $a$.  The gluon fields
have components in the 0 and 3 direction only, so the sum over
the index $\mu$ is restricted to 0 and 3.  The QCD coupling
constant is $g_0$ and $\lambda^{a}$ are the
generators of the
$SU(N_c)$ color group, normalized to
\begin{eqnarray}
&&{\rm trace}\left[\lambda^a\lambda^b\right] =
2\,\delta^{ab}\nonumber\\
&&\sum_{a}\lambda^a\lambda^a=
\frac{2(N_c^2-1)}{N_c}\;\openone \, ,
\label{1eq2aa}
\end{eqnarray}
where matrix multiplication of the $\lambda$ matrices is implied.
As
$N_c\to\infty$, finite results are obtained if
$g_0\to0$ as 
\begin{equation}
g_0=\frac{g}{\sqrt{2N_c}}\, .
\label{1eq2a}
\end{equation} 
where the effective coupling $g$ is a constant. 

While the space-time coordinates, momenta, and Lorentz vector sums in Eq.~(\ref{1eq1}) are restricted
to 1+1 dimension,  we assume that the gamma  matrices have the
usual $4\times4$ Dirac structure.  The quark fields
are therefore a direct product of four Dirac dimensions and $N_c$
color dimensions.   

Note that the action (\ref{A2}) is Lorentz invariant
under the subgroup $G_3$ of Lorentz transformations that leave
the
$xy$ plane invariant.  Specifically, the group is generated by
the hamiltoninian
${\cal H}$, and the boost ${\cal K}_3$, momentum
${\cal P}_3$, and angular momentum ${\cal J}_3$, operators
that generate boosts and translations along the $z$ axis, and
rotations about the $z$ axis.  Four of the six commutation
relations between these generators are zero, and the other
two close 
\begin{eqnarray}
[{\cal K}_3,{\cal P}_3]=-i{\cal H} \qquad 
[{\cal K}_3,{\cal H}]=-i{\cal P}_3  \,,
\label{Lgroup}
\end{eqnarray} 
guaranteeing that $G_3$ is a group.  

Following 't Hooft \cite{thooft} we  introduce light cone
variables:
\begin{eqnarray}
b_{+}&=&\frac{1}{\sqrt{2}}\left(b^0+b^3\right) \\ \nonumber
b_{-}&=&\frac{1}{\sqrt{2}}\left(b^0-b^3\right) \nonumber\\
{\bf b}_\perp&=& \{b_1,b_2\}
\label{1eq3}
\end{eqnarray}
for any arbitrary vector $b$ (in the QCD sector, the only
perpendicular components come from matrix elements of
$\gamma_\perp$). Note  that the scalar product of any two vectors
$a$ and $b$ is
\begin{equation}
a_\mu b^\mu= a\cdot b=a_{+}b_{-}+a_{-}b_{+} -{\bf a}_\perp\cdot
{\bf b}_\perp
\end{equation}
The derivatives are defined
\begin{eqnarray}
\partial_{-}=&&\frac{\partial}{\partial x_+}=
\frac{1}{\sqrt{2}}\left(\partial^0-\partial^3\right)=
\frac{1}{\sqrt{2}}\left(\frac{\partial}{\partial x^0}+
\frac{\partial}{\partial x^3}\right), 
\nonumber \\
\partial_{+}=&&\frac{\partial}{\partial x_+}=
\frac{1}{\sqrt{2}}\left(\partial^0+\partial^3\right)=
\frac{1}{\sqrt{2}}\left(\frac{\partial}{\partial x^0}-
\frac{\partial}{\partial x^3}\right) 
\end{eqnarray}
so that the divergence of a two-vector is
\begin{eqnarray}
\partial_0 b^0 +\partial_3 b^3=\partial_- b_+ + \partial_+
b_-\, .
\end{eqnarray}
In the same way, we can define the $+$, $-$, and $\perp$
components  of the $\gamma$ matrices. The anticommutation
relations are all zero except for
\begin{eqnarray}
&&\{ \gamma_+, \gamma_- \}=2 
\nonumber\\
&&\{ \gamma_x, \gamma_x \}=\{ \gamma_y, \gamma_y \}=-2\, .
\end{eqnarray}

Since the gluon fields are confined to 1+1 dimensions, there
is only one nonvanishing component of the gluon field strength
tensor
\begin{equation}
F_{+-}=-F_{-+}=\partial_+A_- -\partial_
-A_+ +ig_0[A_-,A_+]\, .
\end{equation}
The QCD part of the theory is simplified if we choose
the light  cone gauge, where $A_{-}=0$, so that the commutator
contained in the  field tensor $F_{+-}$ disappears and there is
only one nonzero component ($A_+$) of the gluon field.  
The Lagrangian density (\ref{1eq1}) then reduces to
\begin{equation}
{\cal L}_{QCD}=\frac{1}{2} Tr \left[(\partial_- 
A_{+})^2 \right] + \sum_i
\bar{q}_i\left(i \partial_+
\gamma_- +i \partial_- \gamma_+ -g_0 \gamma_{-}A_{+} 
-m_{0i}  \right)q_i\, .
\label{1eq4}
\end{equation}
The equation of motion for this field is then
\begin{equation}
\partial_-^2A^a_+=\left(\frac{\partial}{\partial x_+}\right)^2
A^a_{+}=-g_0\,\sum_i
\bar{q}_i\,\lambda^a\gamma_{-}\, q_i\, .
\label{1eq5}
\end{equation}
The solution of (\ref{1eq5}) is
\begin{equation}
A^a_{+}(x_{+},x_{-})=g_0 \int dy_{+}\;{\cal
G}(x_{+}-y_{+})\,\sum_i
\bar{q}_i (y_{+},x_{-})\, \lambda^a\gamma_{-} \,q_i(y_{+},x_{-})  
\, ,
\label{1eq6}
\end{equation}
where the Green's function ${\cal G}$ is
\begin{equation}
{\cal
G}(x_{+}-y_{+})=-\frac{1}{2}\,|x_{+}-y_{+}|+
c_1(x_{+}-y_{+})+c_2\, .
\label{1eq7}
\end{equation}
The coefficients $c_1$ and $c_2$ cannot be 
determined without knowing the boundary conditions, so
they are free parameters. The
gauge  condition did not eliminate all superfluous degrees of
freedom,  just as the Coulomb gauge, or the Lorentz gauge, do
not determine uniquely the photon  propagator in QED (Gribov
ambiguity). We can therefore set the  coefficients $c_1$ and $c_2$
equal to zero (a specific choice of gauge)
in order to simplify our calculations.  
 
Einhorn discussed the gauge 
issues related to these two 
parameters and showed that the eigenvalues of the two body 
bound state equation are independent of the choice of $c_1$ and
$c_2$. However, the dressed quark mass does depend on the choice
of $c_2$, and this in turn implies 
that the location of the quark mass pole is gauge dependent. 
Since any physically 
meaningful quantity is gauge invariant, we 
can conclude that the location of the mass pole 
is not physically meaningful, and this can happen only if the
quarks are confined, so that free quark states do not exist. This
is an indirect consequence of confinement.  Direct
consequences of confinement will be discussed in the next
subsection when the two-body bound  state equation is discussed.

Equation~(\ref{1eq6}) shows that the
gluonic field is no longer a  dynamical variable, and that
there are therefore no ghosts.

The Fourier transform of the Green's function (\ref{1eq7})
gives us the gluon ``propagator'', or more precisely the 
momentum dependence of the effective quark-quark interaction. 
With $c_1$ and $c_2$ equal to zero, it can be written
\begin{equation}
\tilde{\cal G}(k_-) =\int^\infty_{-\infty} dx_+\,
e^{i (k_-x_+)}\;{\cal G}(x_+)\to
\left\{\frac{1}{k_-^2}-\delta{(k_-)} 
\int_{-\infty}^{\infty} \frac{d \ell_-}{\ell^2_-}\right\}\, .
\label{1eq8}
\end{equation} 
This is a singular operator that is well defined only when it is
part of an integral over $k_-$. The
second term in Eq. (\ref{1eq8}) was introduced  by Gross
and Milana \cite{GROSS1} in a different context.  Its purpose is
to preserve the condition 
\begin{equation}
{\cal G}(0)=0=\int_{-\infty}^{\infty}dk_-\,\tilde{\cal G}(k_-)
\end{equation}
and it also insures that the potential is finite
for any finite value of $x_+$.

The only variables remaining are the quark fields. The
Feynman rule for the (undressed)  quark propagator is
\begin{equation}
-iS_0(k)=\frac{-i}{m_0-k_- \gamma_+ - k_+ \gamma_-
-i\epsilon}=-i\;\frac{m_0+k_-
\gamma_+ + k_+ \gamma_-} { m_0^2-2 k_+ k_- -  i \epsilon}
\label{1eq9}
\end{equation}
and the quark-``gluon'' coupling is 
\begin{equation}
-i {\cal V}^{a}_{qqg}=-ig_0 \,\lambda^a\gamma_-\, ,
\label{1eq10}
\end{equation}
leading to the following result for the exchange of gluons with
momentum $k_-$ between quarks with Dirac indicies 1 and 2
\begin{eqnarray}
-i {\cal V}_{q{q}}&&= \sum_a\left(-i{\cal V}^{a1}_{qqg}\right) 
{\cal O}_{12} \left(-i{\cal V}^{a2}_{qqg}\right)
\left[-i\tilde{\cal G}(k_-)\right]\nonumber\\ &&=  
-ig_0^2\, \frac{2(N_c^2-1)}{N_c}
\gamma^1_{-}{\cal O}_{12}\gamma^2_-\,
\tilde{\cal G}(k_-)
\to -ig^2 \gamma^1_{-}{\cal
O}_{12}\gamma^2_-\,
\tilde{\cal G}(k_-)
\, , \label{1eq11}
\end{eqnarray}
where we used Eq.~(\ref{1eq2aa}) (with matrix multiplication
of the $\lambda$ matrices implied), and took the limit as
$N_c\to\infty$ using the definition (\ref{1eq2a}).

The quark self-energy and the dressed quark
mass are calculated in the next subsection.

\subsection{Dyson-Schwinger Equation for the dressed
quark}

\begin{figure}
\centerline{
\mbox{
   \epsfxsize=4.0in\epsffile{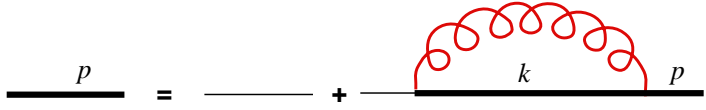}
}
}
\caption{\footnotesize\baselineskip=10ptDyson-Schwinger equation for the quark.  The corkscrew
line is the gluon interaction, the thin line the undressed quark
propagator, and the heavy solid line the dressed quark
propagator.}
\label{ds}
\end{figure}

We determine the dressed single quark propagator,
$S(p)$,  using the (one body) Dyson Schwinger equation (DSE):  
\begin{eqnarray}
S(p)&=&S_0(p)-S(p)\,\Sigma(p)\,S_0(p)\nonumber\\
   &=&S_0(p)- S(p)\left[-ig^2\int\,\frac{d^2k}{(2\pi)^2}\,
\tilde{\cal G}(p_--k_-)\,\gamma_-\, S(k)\,\gamma_-
\right]S_{0}(p)\, ,
\label{DSE}
\end{eqnarray}
shown graphically in Fig. \ref{ds}. The gluon
interaction does not mix quark flavors, and the calculation is
identical for each flavor of quark, so the flavor index is
suppressed. The rainbow approximation (undressed vertices and
the  absence of the quark loops  from the gluon propagator) is
justified  in the large
$N_c$ limit
\cite{thooft}. Since for every internal loop there is a factor
of $\alpha^2=g^2/(2N_c)$, and a multiplicative factor of $\sum_a\lambda^a\lambda^a=2N_c$,
the color dependence disappears. The vertex corrections and the
quark-gluon  vertices do not have a multiplicative factor of
$N_c$, and are therefore  supressed in the large $N_c$ limit. 

%
%

In Eq. (\ref{DSE}), $d^2k=dk_-dk_+$, and since $D$ does 
not depend on $k_+$, it follows 
immediately that the self-energy integral does not depend on
$p_+$ either, and must have the form
$\Sigma(p)=B(p_-)\gamma_-$.  Hence the dressed propagator is
of the form  
\begin{equation}
S(p)=\frac{1}{m_0-p_- \gamma_+ - \left[p_+ - B(p_-) \right]
\gamma_- -i\epsilon}
\, ,
\label{1eq11a}
\end{equation}
where, using $\gamma_-\gamma_+\gamma_-=2\gamma_-$, the
self-energy contribution  is then 
\begin{equation}
B(p_-)=-2i\,g^2\int \frac{dk_-dk_+}{(2 \pi)^2 }      
\frac{k_-\,\tilde{\cal G}(p_- -k_-)}{m_0^2-2k_-(k_+ - B(k_-))-i
\epsilon}\, .
\label{1eq12}
\end{equation}
Performing the $k_+$ integral gives
\begin{equation}
\int dk_+ \frac{k_-}{m_0^2-2k_-(k_+ - B(k_-))-i \epsilon}=
\frac{i \pi}{2} sig(k_-)\, ,
\label{1eq13}
\end{equation}
and substituting this back into Eq. (\ref{1eq12}) gives
\begin{equation}
B(p_-)=\frac{g^2}{4 \pi } \int dk_- \tilde{\cal G}(p_- - k_-)
sig(k_-)\, .
\label{1eq14}
\end{equation}
Using (\ref{1eq8}) for $\tilde{\cal G}(p_- - k_-)$ gives          
\begin{eqnarray}
B(p_-)&=&\frac{g^2}{4 \pi } \int_{-\infty}^\infty
 dk_- \left\{\frac{sig(k_-)}{(p_- - k_-)^2} - 
\frac{sig(p_-)}{(p_- - k_-)^2} \right\}
\nonumber\\
&=& -\frac{g^2}{2\pi p_-}\, .
\label{1eq15}
\end{eqnarray}
Substituting this back into Eq.~(\ref{1eq11}) gives
\begin{eqnarray}
S(p)&=&\frac{1}{m_0-p_- \gamma_+ - \left[p_+
+{\displaystyle\frac{g^2}{2\pi p_-}}\right]
\gamma_- -i\epsilon} =
\frac{m_0+p_-\gamma_+ +\left(p_+
+{\displaystyle\frac{g^2}{2\pi p_-}} \right)
\gamma_-}  {m_0^2-{\displaystyle\frac{g^2}{\pi}}-2 p_+p_-  -i
\epsilon}\, .
\label{1eq16}
\end{eqnarray}
Note that the quark mass pole has been shifted to a {\it
smaller\/} value
\begin{equation}
m_0^2 \rightarrow m^2 = m_0^2 - \frac{g^2}{\pi}\, ,
\label{1eq17}
\end{equation}
where $m$ is the dressed quark mass.  It turns out that
choosing $c_1\ne0$ will not effect the
dressing of the mass, but choosing $c_2\ne0$ would give
\begin{eqnarray}
m^2 = m_0^2 - \frac{g^2}{\pi}\left(1-\pi c_2 |p_-|\right)
\, .
\end{eqnarray}
This gives a mass that is momentum
dependent and not covariant.  Hence the dressed mass is gauge
dependent and unphysical.  With these
cautionary remarks, we choose $c_1=c_2=0$ because the mass is
covariant.  One undesirable feature of this choice is
that $m^2<0$ in the chiral limit. 
This could be corrected by choosing $c_2$ large and positive.

Having obtained the dressed propagator, 
we are able to proceed with the 
two body bound state calculation.


\subsection{Two-body bound states}

\subsubsection{Algebraic form of the two body equations}

\begin{figure}
\centerline{
\mbox{
   \epsfxsize=5.0in\epsffile{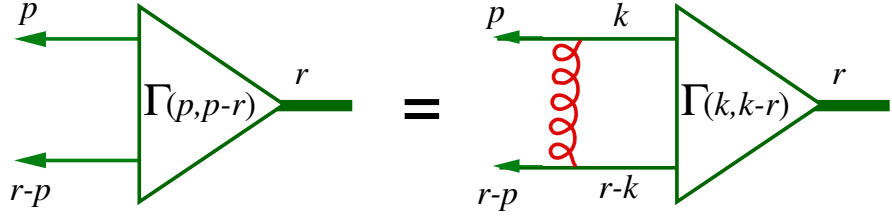}
}
}
\caption{\footnotesize\baselineskip=10ptThe Bethe-Salpeter equation}
\label{bs}
\end{figure}

Consider a bound state of a 
$q\bar q$ pair.
The quark has dressed mass $m_1$ and 
electric charge $e_1$, and 
the antiquark (which might be of a 
different flavor) has 
dressed mass $m_2$ and charge $e_2$. 
The momentum of the bound state is 
$r$, the momentum of the quark is $p$ 
and the momentum of the outgoing antiquark 
is $r-p$, as shown in Fig.~\ref{bs}.  In much of the following
discussion, we will treat the outgoing antiquark as an incoming
quark with momentum
$p-r$, and following this convention label the bound  state
vertex function $\Gamma(p,p-r)$.  For a color neutral state
we may carry out the color sums using (\ref{1eq11}), giving the 
Bethe-Salpeter equation for the bound
state vertex function:
\begin{eqnarray}
\Gamma(p,p-r)&=&i \int 
\frac{d^2 k}{(2 \pi)^2}V(p,k)\, 
\gamma_- \,S_1(k)\, 
\Gamma(k,k-r)\, S_2(k-r) \,\gamma_-\, ,
\label{BSE}
\end{eqnarray}
where $S_1$ and $S_2$ are dressed quark propagators, and the
kernel is the singular operator
\begin{eqnarray}
V(p,k)=g^2\,\tilde{\cal G}(p_- -k_-)\, . \label{kernel2}
\end{eqnarray}
With the substitution 
$\Gamma(p,p-r)=\gamma_-\,G(p,p-r)$ \cite{thooft},
Eq.~(\ref{BSE}) becomes:
\begin{equation}
G(p,p-r)= 4i \int 
\frac{d^2k}{(2 \pi)^2} 
\frac{V(p,k)\,k_- \,(k_--r_-) \,G(k,k-r)}
{\left[m_1^2-k^2 -i\epsilon\right] \left[m_2^2-(k-r)^2
-i\epsilon\right]}\, .
\label{BSE2}
\end{equation}
The formula shows that $G(p,p-r)$ does not depend on $p_+$, and
hence the $k_+$ integration can be carried out immediately.

Integrations of the variable $k_+$ over two quark propagators
occurr frequently, and are worked out in detail in
Appendix \ref{appen:A}.  It is convenient to introduce the
momentum fractions
\begin{equation}
y=\frac{k_-}{r_-}\qquad z=\frac{p_-}{r_-}\, .\label{defz}
\end{equation}
If the fraction $z$ lies in the interval [0,1],
Eq.~(\ref{BSE2}) reduces to
\begin{eqnarray}
\mu_n^2\;\Phi_n(z,r)&=&\left(\frac{\alpha_1+1}
{z}+\frac{\alpha_2+1}{1-z}\right)\,\Phi_n(z,r)- 
\dashint_0^1dy\frac{\Phi_n(y,r)-\Phi_n(z,r)}{(z-y)^2}
\nonumber\\&&\nonumber\\
&\equiv&\frac{\pi}{g^2}\,H(z)\,\Phi_n(z,r) \qquad
\left({\rm if}\;\;z\in[0,1]\right)
\, , \label{phiin}
\end{eqnarray}
where $\dashint$ is the principal value integral, and $\mu_n$,
$\alpha_1$, and $\alpha_2$ are dimensionless parameters
\begin{eqnarray}
\mu_n^2&=&\frac{\pi M_n^2}{g^2}\,
,\qquad \alpha_1=\frac{\pi \,m_{1}^2}{g^2}\, , \qquad
\alpha_2=\frac{\pi\, m_{2}^2}{g^2}\, ,
\label{2eq6}
\end{eqnarray}
with $r^2=M_n^2$ the anticipated mass eigenvalues that will 
emerge from the solution of the equation, and $m_{i}$
the {\it dressed\/} quark masses.  
The two-body ``wave function'' $\Phi_n(z,r)$, is {\it defined\/}
by 
\begin{equation} 
\Phi_n(z,r)= \frac{{\cal N}r_-\,G_n(z,r)}{\Delta(z,M_n^2)}= 
\frac{{\cal N}r_-\,G_n(z,r)}{{\displaystyle
\left(\frac{m^2_1}
{z}+\frac{m^2_2}{1-z}-M_n^2\right)}}=\frac{{\cal N}r_-\,
G_n(z,r)}{{\displaystyle
\frac{g^2}{\pi}\,\left(\frac{\alpha_1}
{z}+\frac{\alpha_2}{1-z}-\mu_n^2\right)}} \, ,
\label{2eq3d}
\end{equation}
where $G_n(z,r)=G(p,p-r)$ is the anticipated eigenfunction
solution for the vertex corresponding to the $n$th bound state,
and  ${\cal N}$ is a normalization constant.
It is chosen so that
\begin{eqnarray}
1=
\int_0^1dz\,\Phi_n^2(z,r)\, ,
\label{norm}
\end{eqnarray}
and will be calculated in subsection \ref{elasticff}.  Note that 
Eq.\ (\ref{phiin}) {\it defines\/} the two-quark Hamiltonian,
$H(z)$, on the interval $0\le z\le1$.  
The {\it bare\/} quark masses enter into the Hamiltonian, while
the {\it dressed\/} quark masses enter into the relation
(\ref{2eq3d}) between the vertex function and the wave function.

If the momentum fraction $z$ lies {\it outside\/} of the interval
$0\le z\le1$, the vertex function defined by Eq. (\ref{BSE2}) is
not zero, but can be obtained by quadrature from the vertex
function defined {\it inside\/} of the interval [0,1].  Using
(\ref{2eqac}) and the definition (\ref{2eq3d}) gives  
\begin{eqnarray}
\Phi_n(z,r)&=&\frac{1}{\displaystyle
\left(\frac{\alpha_1}
{z}+\frac{\alpha_2}{1-z}-\mu_n^2\right)}\, 
\int_0^1dy\frac{\Phi_n(y,r)}{(z-y)^2}\,\quad
\left({\rm if}\;\;z\notin[0,1]\right)
\, .\qquad \label{phiout}
\end{eqnarray}
These definitions and equations will be used in the following
sections when the hadronic currents are computed.

An alternative form of Eq.~(\ref{phiin}) follows if we
integrate the last term by parts:
\begin{eqnarray}
\mu_n^2\;\Phi_n(z,r)&=&\left(\frac{\alpha_1+1}
{z}+\frac{\alpha_2+1}{1-z}\right)\,\Phi_n(z,r)+
\dashint_0^1dy\frac{\Phi'_n(y,r)}{(z-y)}\nonumber\\
&&+\left(\frac{\Phi_n(0,r)-\Phi_n(z,r)}{z}\right)
+\left(\frac{\Phi_n(1,r)-\Phi_n(z,r)}{1-z}\right)
\, , \label{2eqee}
\end{eqnarray}
where $\Phi'_n(y,r)=\partial\,\Phi(y,r)/\partial y$.  If
$\Phi_n(0,r)=\Phi_n(1,r)=0$, which is true in all but the
chiral limit, this form of the equation follows directly from 't
Hooft's principle value prescription
\cite{thooft}, so that our two-body equation is identical with 't
Hooft's, even though we have finite dressed quark masses. 

To avoid confusion later, we call attention to the fact that the
wave function defined by Eqs.\ (\ref{2eq3d}) and (\ref{phiout})
is not identical to the Bethe-Salpeter wave function.  The latter
can be defined 
\begin{eqnarray} 
\Psi(p,r)  &\equiv&
\frac{{\cal N}p_-(p-r)_-\,G(p,p-r)}{(m_1^2-p^2-i\epsilon)
(m_2^2-(p-r)^2-i\epsilon)} 
 \, ,
\label{BSwave}
\end{eqnarray}
where the factor $p_-(p-r)_-$ in the numerator comes from the
$\gamma_-$ factors in the quark propagators $S$.  This
wave function carries the $p_+$ dependence of the quark
propagators.  If we integrate (\ref{BSwave}) over
$p_+$ (using the fact that
$G$ does not depend on $p_+$) we obtain 
\begin{equation} 
\psi(p_-,r)\equiv \int_{-\infty}^\infty dp_+ \Psi(p,r)= \cases{
{\displaystyle{\frac{{\cal N}r_-\,G(p,p-r)}{\Delta(z,r^2)}} =
\Phi(z,r)} & if
$z\in [0,1]$ \cr 0 & if $z\notin [0,1]$ }
 \, .
\label{equalpp}
\end{equation}
This is the ``equal $t_-$'' wave function, and equals the wave
function we are using {\it only in the region\/} $z\in[0,1]$. 
As we have seen, the wave function $\Phi(z,r)$ that solves the 't
Hooft equation is non zero {\it for all\/} $z$.

We conclude this discussion of the two body wave equation by
returning to Eq.\ (\ref{BSE}).  In place of the
replacement $\Gamma(p,p-r)=\gamma_-\,G_n(z,r)$, it could
equally well have been assumed that
\begin{eqnarray}
\Gamma(p,p-r)=\gamma_-\gamma_k\,G_n(z,r)
\, , \label{2eqeee}
\end{eqnarray}
where $k$ is either $x$ or $y$ (recall that we are assuming that the Dirac matrices live in 1+3 dimensional space). 
Note that the Bethe-Salpeter equations with the substitutions
(\ref{2eqeee})  reduce to the same equation (\ref{BSE2}) for
$G_n$, and hence {\it each two-body state is triplely
degererate, with spin structure given by $\gamma_-$,
$\gamma_-\gamma_x$, or $\gamma_-\gamma_y$\/}.   In each of
these three cases the wave functions and bound state mass are
identical.  The additional states with the $\gamma_x$ or
$\gamma_y$ structure are outside the scope of the original 't
Hooft model, but must be considered in reduced QCD.  They are
states with quark spins in the transverse
direction, and will play a major role in the discussion of DIS
below.  


The chiral limit is exactly soluable and of some interest.  If
$m_{0i}=0$, then Eq.\ (\ref{phiin}) has the normalized ground
state solution
\begin{eqnarray}
M_0=0\, ,\qquad \Phi_0(z,r)= 1
\, .
\end{eqnarray}
Outside of the interval
$[0,1]$, Eq.\ (\ref{phiout}) gives the same result, so that
\begin{eqnarray}
\Phi_0(z,r)= 1
\label{chiralsol}
\end{eqnarray}
for all $z$ in $(-\infty,\infty)$.
 

\subsubsection{Confinement in the 't Hooft model}

\begin{figure}
\centerline{
\mbox{
   \epsfxsize=3.0in\epsffile{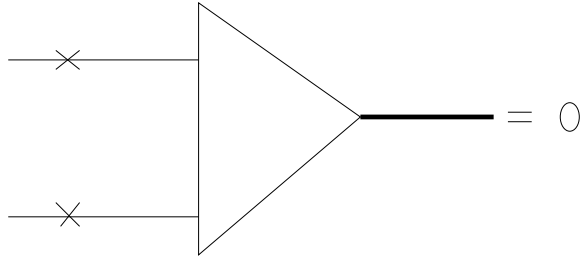}
}
}
\caption{\footnotesize\baselineskip=10pt Diagrammatic representation of confinement via two-body interaction.  The $\times$ indicates that the dressed quark is on-shell, and at the point when both quarks are on-shell, the vertex function is zero.}
\label{confinement}
\end{figure} 

The 't Hooft model illustrates how confinement can be realized
in two (apparently different) ways.  If the singularities of the
confining interaction are left unregularized, as they were in 't
Hooft's original paper \cite{thooft}, the quark masses are
infinite.  Even if the singularity is regularized as in this
paper, whenever
$m_{0i}^2<g^2/\pi$ the dressed quark mass will be unphysical, and
there is no real quark mass pole.  This insures that the quarks
are confined.

In the second case ($m_{0i}^2>g^2/\pi$ and $m_i^2>0$) the dressed quark mass is real and positive and the there is a quark mass pole.  What keeps the
quarks confined in this case?  The answer is that the quark
vertex function $G_n(z,r)$ has a {\it zero\/} at precisely the
value of $z$ where all of the quarks in the bound state could
be on mass shell.  This requirement is illustrated
diagramatically in Fig.\ \ref{confinement}, and seems miraculous.
However, as discussed in Ref.\ \cite{SG}, it is only
a consequence of fact that the {\it wave function\/} is finite at
the pole (because the wave equation permits no singularities),
and therefore the vertex function 
\begin{eqnarray}
G_n(z,r)=\left[\frac{m_1^2}{z} +\frac{m_2^2}{1-z}-M_n^2 \right]
\frac{\Phi(z,r)}{{\cal N} r_-}
\end{eqnarray}
must be zero.  Precisely the same mechanism works in the
Schr\"odinger description of two nonrelativistic particles bound
by a linear (or any other) confining potential. 

In summary, quark mass poles are always unphysical because 
the dressed mass is gauge dependent, and bound state poles are 
physical because they are gauge invariant.  In
this respect the undressed quark mass is the only physically
significant parameter.  The two ways of viewing confinement are
therefore not exclusive,  but equivalent, since they are related
through gauge transformations.


\subsubsection{Numerical solutions for meson wave functions}

The bound state equation was solved numerically using modified
cubic splines.  The splines and the technique are fully described
in Appendix \ref{appen:splines}.  Each spline is a smooth
function (continuous up to its third derivative) and has support
over a small region of the $z$ axis.  The splines overlap, 
so that by independently adjusting the height of each spline it
is possible to smoothly approximate any function defined on
the interval $[0,1]$.  


\begin{table}
\caption{Numerical values of $M_n^2$ for $m_{01}=m_{02}=1.5$ and
various $n_s$ (number of splines) and $n_g$ (number of guass points for each spline segment).}
\begin{tabular}{ r r c c  c c c c  c c c c c c}
 level  & $\quad n_s$ & $\;$ & \multicolumn{3}{c}{$20$} &
$\quad$ &
\multicolumn{3}{c}{$40$} & $\quad$ &  
\multicolumn{3}{c}{$60$}\\
$n$ & $n_g$ &  & 2 & 4 & 6 & & 2 & 4 & 
6 & & 2 & 4 & 6 \\
\colrule
 1 & & & 13.730& 13.732 & 13.732 & & 13.717 & 13.718 & 13.718 &
& 13.714 & 13.715 & 13.715\\
 2 & & & 25.613 & 25.614 & 25.615 & & 25.572 & 25.573 &
25.573 & & 25.563 & 25.564 & 25.564\\
 10 & & & 108.95 & 109.35 & 109.36 & & 108.42 & 108.43 &
108.43 & & 108.28 & 108.28 & 108.28\\
 20 & & & 186.17 & 210.60 & 210.58 & & 208.82 & 209.69
& 209.69 & & 208.88  & 208.97 & 208.97 \\
 30 & & & --& -- &-- & & 299.64 & 314.02  & 314.02 & & 308.04 &
309.38  & 309.38 \\
 40 & & & --& -- & -- & & 362.04 & 409.77  &  409.74 & &  410.31
 & 411.24 & 411.24 \\ 
\end{tabular}
\label{tab:2}
\end{table}
 

\begin{table}
\caption{Numerical values of $M_n^2$ for $m_{01}=m_{02}=0.5$ and
various $n_s$ and $n_g$.}
\begin{tabular}{ r r c c  c c c c  c c c c c c}
 level  & $\quad n_s$ & $\;$ & \multicolumn{3}{c}{$20$} &
$\quad$ &
\multicolumn{3}{c}{$40$} & $\quad$ &  
\multicolumn{3}{c}{$60$}\\
 & $n_g$ &  & $4$ & 6 & 8 & & $4$ & 6 & 
8 & & 4 & 6 & 8 \\
\colrule
 1 & & & 2.6196 & 2.6625 & 2.6841 & & 2.6485 & 2.6786 & 2.6936 &
& 2.6616 & 2.6858 & 2.6979\\
 2 & & & 10.540 & 10.650 & 10.706 & & 10.616 & 10.692 &
10.731 & & 10.648 & 10.710 & 10.741\\
 10 & & & 86.170 & 86.371 & 86.477 & & 86.055 & 86.288 &
86.409 & & 86.197 & 86.400 & 86.503\\
 20 & & &  199.44 & 199.43 & 199.43 & & 183.93 & 184.11
& 184.21 & & 183.53  & 183.77 & 183.90 \\
 30 & & & --& -- &-- & & 287.16 & 287.22  & 287.25 & & 282.09 &
282.27  & 282.36 \\
 40 & & & --& -- & -- & & 399.18 & 399.15  &  399.15 & &  383.13
 & 383.22 & 383.27 \\ 
\end{tabular}
\label{tab:1}
\end{table}

We have modified the standard spline technique to insure that the
numerical solutions satisfy the boundary conditions exactly. 
These boundary conditions are determined by examing
Eq.~(\ref{phiin}) in the vicinity of $z\sim0$ and $z\sim1$
\cite{thooft,gl}, and require the wave function go like 
\begin{eqnarray}
&\Phi_n(z,r) \to z^{\beta_1}\quad\quad\;&\quad({\rm
as}\;\;z\to0)\quad{\rm with}\;\;\pi\beta_1=-\alpha_1\,\tan
\,(\pi\beta_1)\nonumber\\ 
&\Phi_n(z,r) \to
(1-z)^{\beta_2}&\quad({\rm as}\;\;z\to1)\quad{\rm
with}\;\;\pi\beta_2=-\alpha_2\,\tan
\,(\pi\beta_2)\, . \label{betas}
\end{eqnarray}
Note that the $\beta_i=0$ in the chiral limit (when
$\alpha_i=-1$),  consistent with the exact solution
(\ref{chiralsol}).   To fit these boundary conditions the splines
closest to $z=0$ and $z=1$ have these fractional powers built
into their functional form, as described in Appendix
\ref{appen:splines}.

The wave function is expanded into $n_s$ splines, each with four
segments (except for those at the boundary, which have only
three). The splines overlap, and taking into account that there
must be at least three splines (one regular spline in the center
and one modifed spline at each end), the total number of
segments is $n_s+1$, dividing the interval into segments of length
$1/(n_s+1)$.  The equation is turned into a matrix equation by
integrating over each segment using $n_g$ gaussian points. 
Tables \ref{tab:2} and \ref{tab:1} show the numerical stability. 
If the quark mass is larger than unity, better than 1\% accuracy
is achieved with only two gauss points per interval, provided
$n\lesssim n_s$.  As the level number $n$ increases above $\sim
n_s/2$ toward $n_s$, the reliability of the calculation
decreases.  To study highly excited states, it is sufficient to
use $n_g=2$, and push $n_s$ as high as possible.

The situation is somewhat different if the bare quark masses are
less than unity (Table \ref{tab:1}).  In this case the wave
function is very steep at the boundaries
and 4 or 6 gaussian points per interval are needed to get 1\%
accuracy for the low lying states. In a subsequent paper \cite{BGII}, we will limit our numerical discussion of duality to cases with $m_{0i}>1$ where the
solutions are very stable, and there are no poles in the region
outside of [0,1] [see the brief remarks following Eq.\
(\ref{Rlimit})].

\begin{figure}[hbt]
\centerline{
\mbox{
   \epsfxsize=3.0in\epsffile{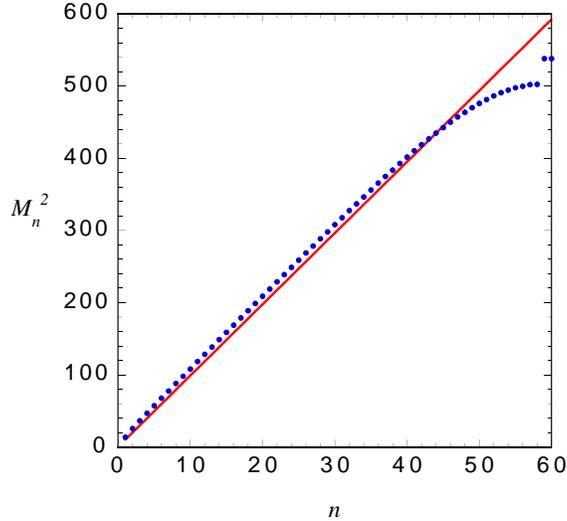} 
}
}
\caption{\footnotesize\baselineskip=10ptThe square of the bound state mass (in units of
$g^2/\pi$) versus the order of the state $n$, computed using 60
splines.  The straight line is the linear relation $n\,\pi g^2$. 
Note that the masses follow this linear relation up to near
$n\sim n_s =60$, where the departure from linearity is due to
numerical inaccuracy. }
\label{mass}
\end{figure} 

\begin{figure}[hbt]
\centerline{
\mbox{
   \epsfxsize=4.0in\epsffile{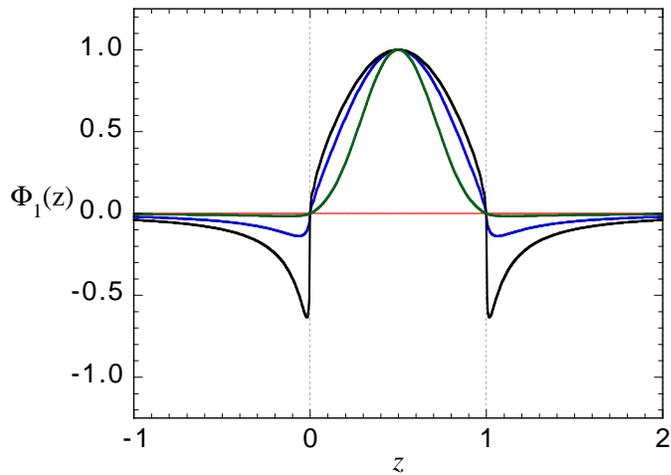}
}
}
\caption{\footnotesize\baselineskip=10ptThe ground state wave functions for equal bare quark
masses $m_0$ of 1.1, 1.5, and 3 (the wave functions become
steeper at $z=0$ and $z=1$ as the quark
mass decreases).   In all cases 60 splines were used, and the
wave function was normalized to a maximum value of unity.}
\label{1st}
\end{figure} 
\begin{figure}[hbt]
\centerline{
\mbox{
   \epsfxsize=4.0in\epsffile{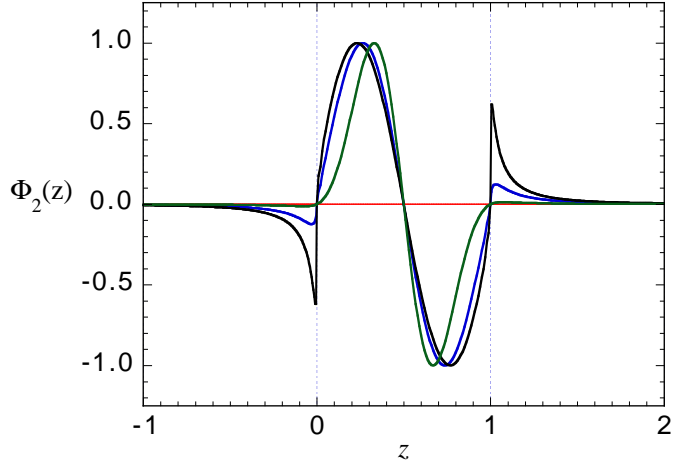} 
}
}
\caption{\footnotesize\baselineskip=10ptThe second state (first excited state) 
for the same three quark masses shown in Fig.~\ref{1st}.  (See
caption to Fig.~\ref{1st}.)}
\label{2nd}
\end{figure} 
\begin{figure}[hbt]
\centerline{
\mbox{
  \epsfxsize=4.0in\epsffile{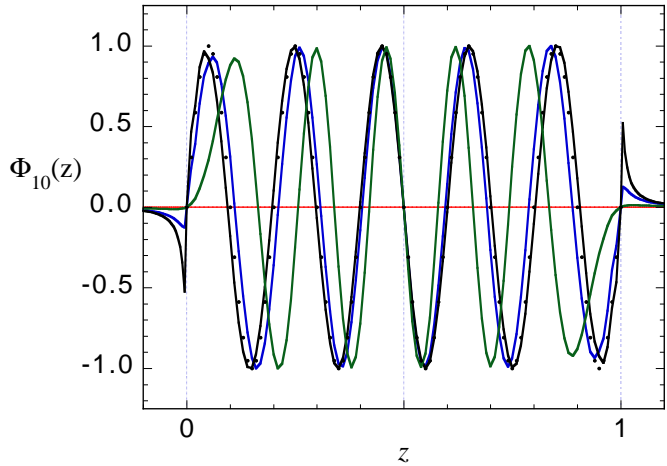}
}
}
\caption{\footnotesize\baselineskip=10ptThe tenth state for the same three quark masses 
shown in Fig.~\ref{1st}.  (See caption to Fig.~\ref{1st}.)  In
this figure the dots are $\sin(10\pi z)$, showing that this is an
excellent approximation for small quark masses.}
\label{10th}
\end{figure} 

\begin{figure}[hbt]
\centerline{
\mbox{
   \epsfxsize=6.0in\epsffile{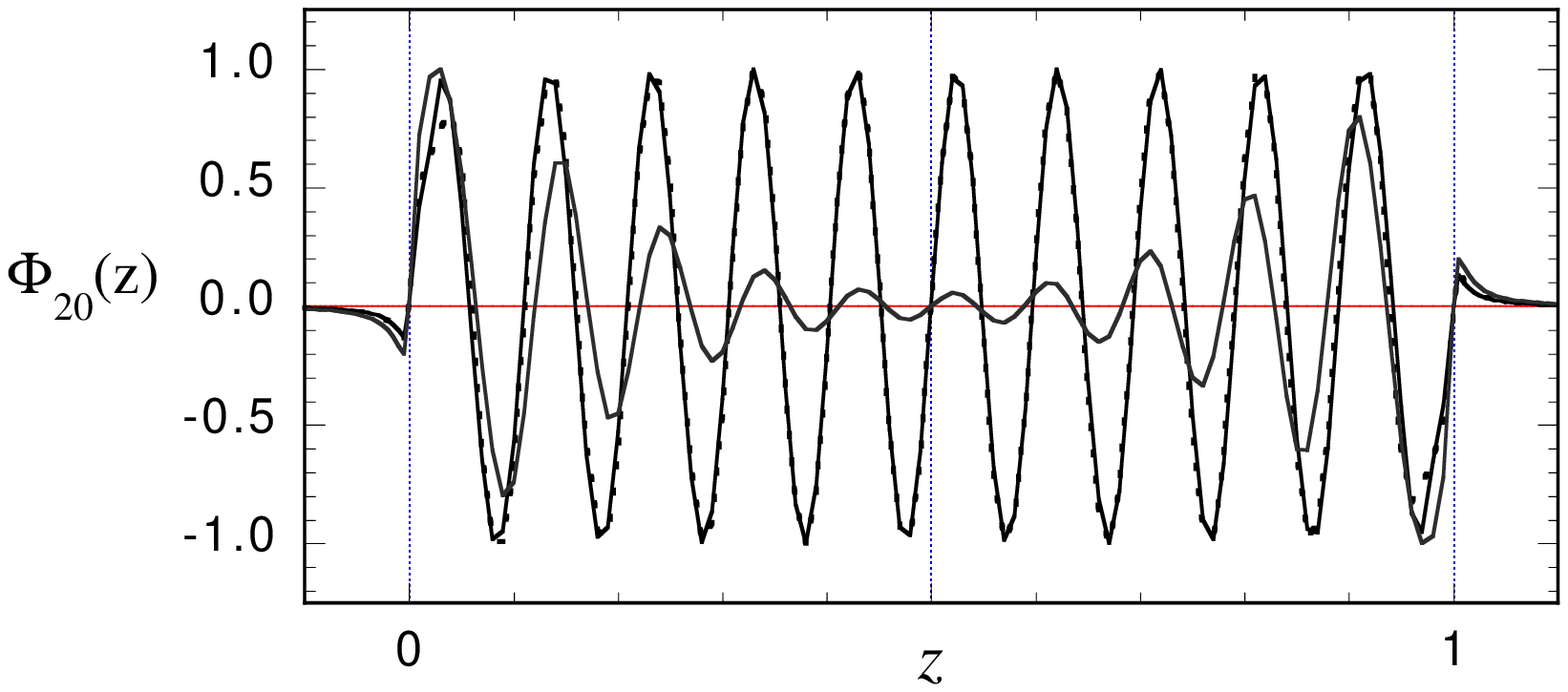}
}
}
\caption{\footnotesize\baselineskip=10ptThe 20th state for and equal quark mass of 1.5.  The
small amplitude curve was calculated for $n_s=20$, and completely
fails to represent the state. The dotted and full amplitude
solid lines are calculated for $n_s = 40$ and 60,
respectively.  Note that $n_s=40$ does well except for the first
and last oscillation.}
\label{20th}
\end{figure} 

Figure \ref{mass} shows the linearly rising Regge trajectory
characteristic of the 't Hooft model. Figures \ref{1st}-\ref{10th}
show the ground, first excited and 10th level for three cases with equal
quark masses of different values.  Note the ``tails'' of the wave functions outside of the
region [0,1], given by Eq.\ (\ref{phiout}).  These tails are
large for small quark masses, but shrink quickly as the quark
mass increases, and are also small for highly excited states. 
Figure \ref{20th} shows the 20th level for spline numbers 20, 40,
and 60.  Note that the state is not fully described even with 40
splines, supporting the observation that an accurate description
of states of order $n$ requires a spline number of approximately
$\sim 2n$.

\subsection{Scattering of quarks and antiquarks in the presence
of confinement}

\subsubsection{Completeness relation and the two-body Greens
function}

The Hamiltonian defined in Eq.~(\ref{phiin}) is hermitian on
the interval $[0,1]$ and it is straightforward to show that
the different solutions are orthogonal on this interval. 
With the normalization given in Eq.~(\ref{norm}) 
\begin{equation}
\int_0^1dz\,\Phi_n(z,r_n)\Phi_m(z,r_m) = \delta_{nm}\, .
\end{equation}
The completeness of the eigenfunctions implies that
\begin{equation}
\sum_n\,\Phi_n(z,r_n)\Phi_n(z',r_n) = \delta(z-z')\qquad{\rm
if}\; z,z'\in [0,1]\, . \label{complete}
\end{equation}
The Greens function is therefore
\begin{equation}
G(z,z',q^2)=\sum_n\,\frac{
\Phi_n(z,r_n)\Phi_n(z',r_n)}
{q^2-M_n^2}
\, ,
\end{equation}
where $q$ is any external momentum.  The completness relation
shows that $G$ has the property of a Greens function
\begin{eqnarray}
\left\{q^2-H(z)\right\}G(z,z',q^2)&=&\sum_n\,
\left(q^2-H(z)\right)\frac{
\Phi_n(z,r_n)
\Phi_n(z',r_n)} {q^2-M_n^2}\nonumber\\
&=& \delta(z-z')\qquad{\rm
if}\; z,z'\in [0,1]
\, .
\end{eqnarray}
The same argument also gives
\begin{eqnarray}
G(z,z',q^2)\,\left\{q^2-H(z')\right\}&=&\delta(z-z')\qquad{\rm
if}\; z,z'\in [0,1]
\, .
\end{eqnarray}
%

\subsubsection{The scattering matrix}
In preparation for the study of electromagnetic interactions, it
is useful to introduce the $q\bar q$ scattering matrix, defined by
the following infinite series 
\begin{eqnarray}
&&M(p',p;q)\,\gamma_{1-}\otimes\gamma_{2-}
=V(p',p)\,\gamma_{1-}\otimes
\gamma_{2-}\nonumber\\
&&\qquad+\,i\int\frac{d^2k}{(2\pi)^2}
V(p',k)\,V(k,p)\,\left[\gamma_{-}S_1(k)\gamma_-\right]_1
\otimes\left[\gamma_{-}S_2(k-q)\gamma_-\right]_2
\nonumber\\
&&\qquad+i^2\int\frac{d^2k'}{(2\pi)^2}\int\frac{d^2k}{(2\pi)^2}
V(p',k')\,V(k',k)\,V(k,p)\,\nonumber\\
&&\qquad\qquad\times
\left[\gamma_-S_1(k')\gamma_{-}S_1(k)\gamma_-\right]_1
\otimes\left[\gamma_{-}S_2(k-q)\gamma_-S_2(k'-q)\gamma_-\right]_2 
\nonumber\\
&&\qquad +\cdots \, .\quad \label{eq248}
\end{eqnarray}
The momentum are defined as for the two-body bound state,
except that here $q$ is the unconstrained momentum of the
$q\bar q$ pair and $q^2$ is not an eigenvalue.  The kernel
$V$ is the singular operator (\ref{kernel2}). 
The singularities of this operator will never be exposed because
we will limit application of this scattering equation to cases
where we integrate over either the initial or final momentum.
Note that the structure of the equation anticipates the
reduction of the dressed propagators
\begin{eqnarray}
[\gamma_-\,S_i(k)\,\gamma_-]_i=\frac{k_-\,
[\gamma_-\gamma_+\gamma_-]_i}
{m_i^2-2k_-k_+}=\frac{2k_-}{m_i^2-2k_-k_+}\,\gamma_{i-}\, ,
\end{eqnarray}
so that the factor of $\gamma_{1-}\otimes \gamma_{2-}$ is
common, and the series (\ref{eq248}) reduces to
\begin{eqnarray}
M(p',p;q)
=V(p',p)&+&\,i\int\frac{d^2k}{\pi^2}\;
\frac{V(p',k)\,V(k,p)}
{d_2(k_-,k_+,q)}\nonumber\\
&+&i^2\int\frac{d^2k'}{\pi^2}\int\frac{d^2k}{\pi^2}
\;\frac{V(p',k')\,V(k',k)\,V(k,p)}
{d_2(k'_-,k'_+,q)\,d_2(k_-,k_+,q)}+\cdots\, ,\qquad\quad
\label{eq250}
\end{eqnarray}
where $d_2$ was defined in Eq.~(\ref{d2}).
From this result we can conclude that $M$ does not depend on
$p'_+$ and $p_+$.

In applications below, we will encounter this scattering
series in the following form
\begin{eqnarray}
\left<MG{\cal O}\right>(p';q)&=&\int\frac{d^2p}
{\pi^2}\frac{{M}(p',p;q)\,{\cal
O}(p_-,q)}{d_2(p_-,p_+,q)}
=\int\,\frac{d^2p}
{\pi^2}\frac{{V}(p',p)\,{\cal
O}(p_-,q)}{d_2(p_-,p_+,q)}
\nonumber\\
&&+i\int\frac{d^2k}
{\pi^2}
\frac{V(p',k)}
{d_2(k_-,k_+,q)}\int\frac{d^2p}
{\pi^2}\frac{{V}(k,p)\,{\cal
O}(p_-,q)}{d_2(p_-,p_+,q)}
\nonumber\\ 
&&+i^2\int\frac{d^2k'}{\pi^2}\int\frac{d^2k}{\pi^2}
\;\frac{V(p',k')\,V(k',k)}
{d_2(k'_-,k'_+,q)\,d_2(k_-,k_+,q)}\int\frac{d^2p}
{\pi^2}\frac{{V}(k,p)\,{\cal
O}(p_-,q)}{d_2(p_-,p_+,q)}
\, ,\quad\quad\quad \label{eq249}
\end{eqnarray}
where ${\cal O}(p_-,q)$ is an operator that, by assumption,
does not depend on $p_+$.  In Appendix \ref{appen:B} we show
how to use the Greens function to write this series as
\begin{eqnarray}
\int_0^1dz\,{\rm M}(z',z;q^2)\,\frac{{\rm
O}(z,q^2)}{\Delta(z,q^2)}&=&
\sum_n\Delta(z',M_n^2)\,
\frac{\Phi_n(z',r_n)}{M_n^2-q^2}
\,\int_0^1dz\,\Phi_n(z,r_n) \,{\rm O}(z,q^2)
\, ,\qquad \label{eq267}
\end{eqnarray}
where we replaced
$(q_-)^2M(p',p;q)/\pi\to{\rm M}(z',z;q^2)$ and ${\cal
O}(p_-,q)\to{\rm O}(z,q^2)$.    

This equation has a nice physical interpretation.  It displays
the ``scattering amplitude'' as the sum over the propagation of
the bound states, which is all confinement will allow (without
meson decay mechanisms, which are ignored in this paper, no
cuts are possible).  The only singularities in $q^2$ that
can occur are poles at the bound state masses $M_n^2$.
 
With these tools in place we are able to 
study the electromagnetic interaction of hadrons.


\section{Reduced QCD and the electromagnetic sector}
\label{RQCD}


We return to the discussion of the Lagrangian density introduced
in Eq.\ (\ref{A2}) and look at the electromagnetic sector. 
In rQCD the electromagnetic fields (and the electron currents) are
extended to 1+3 dimensions.  The Lagrangian density is therefore
\begin{eqnarray}
{\cal L}_{QED}(t,{\bf r}) = -\frac{1}{4} F_0^{\mu
\nu}(t,{\bf r})F_{0\,\mu \nu}(t,{\bf r}) &+& \sum_i e_i\;
\bar{q}_i(t,z)\gamma_{\mu} q_i(t,z)\;A^{\mu}_0(t,{\bf r})
\nonumber\\
&-&e\,J^e_\mu(t,{\bf r})\;A^{\mu}_0(t,{\bf r})   \, ,
\end{eqnarray}
where $A_0(t,{\bf r})$ is the electromagnetic field (with the subscript 0 distinguishing it from the gluon field), the vector
sums are now over 4 dimensions with $\mu=\{0,1,2,3\}$, $e_i$ is
the electromagnetic charge of quarks with flavor $i$,
$J^e_\mu(t,{\bf r})$ is the electron current that produces the
electromagnetic field, and 
\begin{eqnarray}
F^{\mu \nu}_0=&&\partial^{\mu}{\cal A}^{\nu}_0-\partial^{\nu}
{\cal A}^{\mu}_0  \,,
\label{1eq2qed}
\end{eqnarray} 
is the electromagnetic field tensor.  Because the electron
current exists in four dimensions, all four components of the
electromagnetic field will be non-zero, in general, and all
four components of the quark current will be excited by
electromagnetic scattering.


We now apply this discussion to electron scattering, where
$Q^2=-q^2>0$, so that $q_-<0$.  The initial particle (quark or
hadron) has momentum
$p=p'-q$ and the final particle momentum $p'$, with momentum
fractions 
\begin{eqnarray}
z'=\frac{p'_-}{q_-}\qquad\frac{p_-}{q_-}=z'-1\, .\qquad\qquad
\label{frac1}
\end{eqnarray}
If the final particle is on shell, then $p'_->0$ and $z'<0$, and
$z'\to-\infty$ as $Q^2\to0$.  We will use the notation $j^\mu_0$
to denote the {\it bare\/} quark current {\it
operator\/}, which is
\begin{eqnarray}
j^\mu_0=\gamma^\mu \, .
\end{eqnarray}

While the dressing of the strong quark-gluon 
vertex is supressed by the large $N_c$ limit \cite{thooft}, the
quark electromagnetic vertex can be dressed by gluon exchanges.
The longitudinal part of the dressed electromagnetic vertex
($j^{\mu}$) can be computed directly from the Ward-Takahashi
identity
\begin{equation}
q_{\mu}\,j^{\mu}(p',p)=S^{-1}(p)-S^{-1}(p')\, ,
\end{equation}
where $p+q=p'$. Using the dressed propagator (\ref{1eq16}),
we find that
\begin{eqnarray}
j_-=&&\gamma_- \nonumber \\
j_+=&&\gamma_+ - \gamma_-\,\frac{g^2}{2
\pi}\frac{1}{p'_{-}p_{-}} = \gamma_+ +\gamma_-\, \frac{g^2}{2
\pi}\frac{1}{q_-^2\,z'(1-z')}
\, . \label{current1}
\end{eqnarray}
Consequently the $j_-$ component is unmodified. 

\subsection{Electromagnetic coupling to quarks and the quark
form factor} 
\label{quarkem}

\begin{figure}
\centerline{
\mbox{
   \epsfxsize=6.0in\epsffile{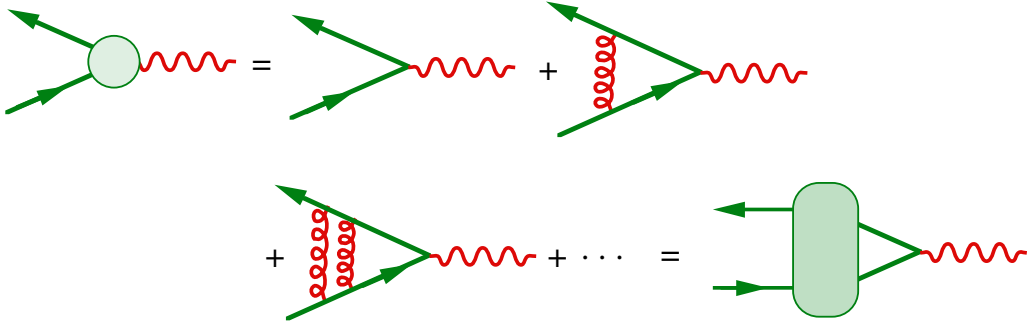}
}
}
\caption{\footnotesize\baselineskip=10pt Diagrammatic picture of the gluonic dressing of the quark current.}
\label{quarkcurrent}
\end{figure} 
 
It is instructive to obtain the full quark current directly
from the dressing of the quark-photon vertex.  The vertex is
dressed by successive gluon exchanges, as shown
diagramatically in Fig.~\ref{quarkcurrent}.  When coupling to a
photon, there is no flavor change, and hence, for each flavor
$i$, this series is
\begin{eqnarray}
&&j_i^\mu(p',p'-q)=\gamma^\mu +i
\int\frac{d^2k}{(2\pi)^2}\,V(p',k)\,
\gamma_-\,S_i(k)\,\gamma^\mu\,S_i(k-q)\,\gamma_-\nonumber\\
&&\quad+\,i^2\int \frac{d^2p}{(2\pi)^2}\int
\frac{d^2k}{(2\pi)^2}\,V(p',p)\,V(p,k)\,
\gamma_-\,S_i(p)\,\gamma_-\,S_i(k)\,\gamma^\mu
\,S_i(k-q)\,\gamma_-\,S_i(p-q)\,\gamma_-\nonumber\\
&&\quad+\cdots\, .
\label{current1a}
\end{eqnarray}

This series is evaluated in Appendix \ref{appen:C} using the
methods developed in Appendix \ref{appen:B}.  The final results
are conveniently expressed in term of a quark form factor,
defined to be
\begin{eqnarray} 
F_i(z',Q^2)&\equiv&   
\sum_n\Delta_{i}(z',M_n^2)
\frac{\Phi_n(z',r_n)}{M_n^2+Q^2}\;
\int_0^1dy\,\Phi_n(y,r_n)\, ,
\label{ff1}
\end{eqnarray}
where  $z'=p'_-/q_-$ and $Q^2=-q^2$ in anticipatation of
applications to electron scattering.  Note that the form factor
is expressed as a sum of bound state poles at $Q^2=-M_n^2$,
arising from the propagation of the bound states that couple to
the photon. Vector dominance is a rigorous consequence of this
model.

Using this form factor, the currents reduce to 
\begin{eqnarray} 
j_{i-}(p',q)&=&\gamma_- + \gamma_-\,F_i(z',Q^2) \label{ff3aa}\\
j_{i+}(p',q)&=&\gamma_+ -
\gamma_-\,\left\{
\frac{g^2}{2\pi}\frac{1}{p'_-p_-}  +
\frac{q_+}{q_-}\,F_i(z',Q^2)\right\}
\label{ff4} \\
j_i^x(z',q^2)&=&\gamma_x  -
\gamma_x\,\gamma_-\,\left\{\frac{q_-}{m_{0i}}\,\frac{g^2}{2\pi}
\frac{1}{p'_-p_-}+\frac{q_+}{m_{0i}}\,F_i(z',Q^2)
\right\}
\, .\qquad \label{ff6}
\end{eqnarray}
Note that the terms independent of the quark form factor $F_i$
are identical to the expected result (\ref{current1}), and that
the contribution from the form factor is {\it purely
transverse\/}, in that    
\begin{eqnarray} 
\Bigl[q_+j_{i-}(p,q)+q_-j_{i+}(p,q)\Bigr]\Big|_{F_i\;{\rm
term}} &=&\gamma_-
\,F_i(z,Q^2)\,\left\{ q_+-q_-\frac{q_+}{q_-}\right\}=0\, .\qquad
\label{ff5}
\end{eqnarray}
This result mimics the method for insuring current
conservation developed in Ref.~\cite{RG}, and gives some
evidence that that method is dynamically sound. 

These results are used in the next subsection 
to calculate the hadronic transition current.

\subsection{The transition current}
\label{transff}

\begin{figure}
\centerline
\mbox{
{\vspace*{-0.5in}
   \epsfxsize=6.0in\epsffile{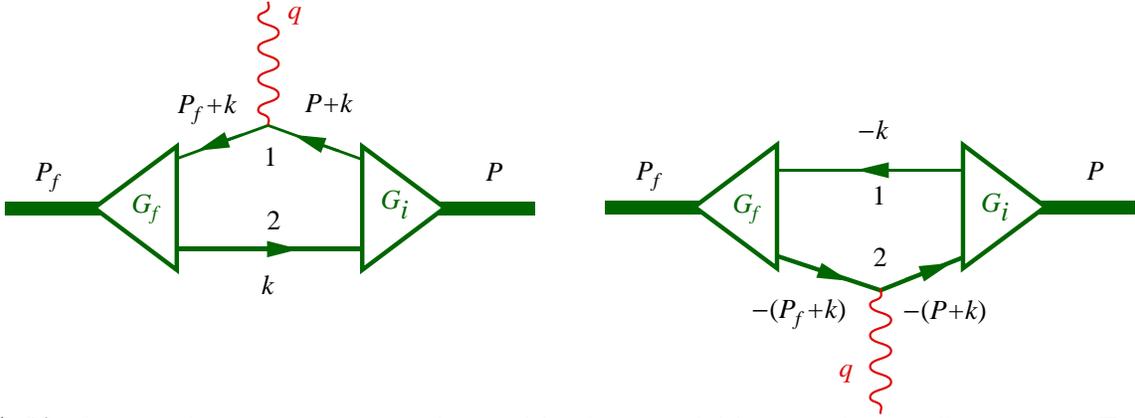}
}
}
\caption{\footnotesize\baselineskip=10pt Currents for the electromagnetic transition from the
initial state $i$ to the final state $f$.  The left hand diagram
is the contribution from the electromagnetic interaction with the
quark with mass $m_1$ and charge $e_1$ and the right hand diagram
is the same for the antiquark with mass $m_2$ and charge $e_2$.}
\label{hadronicamplitude}
\end{figure}

In the presence of confinement, and without any meson decay
mechanisms, the quarks in a 
$q\bar q$ bound state must remain bound, even after they absorb
an energetic photon.  In this section we calculate the
transition current for the process $\gamma^*+M_i\to M_f$, where
$M_f$ is the mass of a (possibly excited) final bound state and
$M_i$ is the mass of the initial (ground) state.   

The transition current consists of the sum of two
contributions:  one in which the photon is absorbed by the
quark with dressed mass $m_1$, and a second in which the
photon is absorbed by the antiquark with mass $m_2$. The
momentum used to label each of these processes are shown in
Fig.~\ref{hadronicamplitude}; the initial bound state has
momentum $P$ and the final state has momentum $P_f=P+q$.  The
transition current coming from the first term is
\begin{eqnarray}
&&\left<f|{\cal J}^\mu(P_f,P)|i\right>\Big|_{e_1{\rm
term}}=ie_1N_c\int\frac{d^2k}{(2
\pi)^2} G_f\left(-P_f-k,-k\right)
G_i\left(P+k,k \right)  \nonumber\\
&&\qquad\qquad\qquad\qquad\times Tr \left[ 
\gamma_-\left\{\begin{array}{c}1\\ \gamma_x\end{array}
\right\} S_1(P_f+k) j_1^\mu S_1(P+k) 
\gamma_- S_2(k) \right]\nonumber\\
&&\qquad\quad=i \int\frac{d^2k}{(2
\pi)^2}\frac{ e_1N_c\,G_f\left(-P_f-k,-k\right)
G_i\left(P+k,k \right)\,N^\mu }{[m_1^2-(P_f+k)^2-i\epsilon]
[m_1^2-(P+k)^2-i\epsilon] [m_2^2-k^2-i\epsilon]}
\, ,\qquad
\label{3eq1-}
\end{eqnarray}
where $j_1^\mu$ is the quark current operator for quark 1, as
worked out in subsection~\ref{quarkem}, $N^\mu$ is the spin 
dependent numerator discussed below, and we have allowed for the
possibility that the final state may have {\it either\/} the
$\gamma_-$ {\it or\/} the $\gamma_-\gamma_x$ structure discussed
above (we will not consider the $\gamma_-\gamma_y$ states; they
are needed only if we consider the $y$ component of the
transverse current and give results identical to the results
for the $x$ components).  Note the factor of $N_c$ coming from
the sum over all possible quark colors.
 
Only the term in the numerator of the 
$S_2$ propagator equal to
$k_-\gamma_+$ can give a nonzero result, but the terms that
contribute from the $S_1$ propagators depend on the matrix form
of the final state, and of $j_1^\mu$.  Because the trace of
a product of gamma matrices with only one factor of $\gamma_x$
or
$\gamma_y$ is zero, the  transverse currents (with $\gamma_x$ or
$\gamma_y$) will be zero unless the final
state also includes $\gamma_x$ (or $\gamma_y$ for the $y$
component of the current).  Hence we have the rules:
(i) $x$-type transverse currents couple $\gamma_-$ states to
$\gamma_-\gamma_x$ states only, and (ii)
longitudinal currents couple $\gamma_-$ states to each
other.
Because of the exact degeneracy of these states, the results in
both of these cases use the same momentum-space wave functions.

The integral (\ref{3eq1-}) is evaluated in Appendix
\ref{appen:D}.  The result for the $-$ component of the current
is  
\begin{eqnarray}
\left<f_-|{\cal J}_{-}(P_f,P)|i_-\right> \Big|_{e_1\,{\rm
term}}&=&\frac{16 e_1N_c\,P_{-}}{{(2
\pi)\cal N}^2} \left\{
\int_{y}^1  + 
\int_{0}^{y} {\cal R}\right\} d\xi
\;\Phi_f(\xi',P_f)\Phi_i(\xi,P) \nonumber\\
&&\qquad\qquad\times\left[1+F_1(\eta,Q^2)\right]
\, ,\qquad\qquad\qquad\qquad
\label{3eq1b}
\end{eqnarray}
where 
\begin{eqnarray}
y=-\frac{q_-}{P_-}\, ,\qquad\xi'=\frac{\xi-y}{1-y}\, , \qquad
\eta=\frac{y-\xi}{y} \label{fraccur}
\end{eqnarray}
[for definitions of all the momentum fractions see 
Eqs.~(\ref{frac2}) and (\ref{frac3})], the state $f_-$ has the
same structure as the initial state (assumed to be $\gamma_-$),   
$F_1$ is the quark form factor defined in Eq.~(\ref{ff1}) and
${\cal R}$ is given in Eq.~(\ref{Rlimit}).  The plus and
transverse components of the current are  
\begin{eqnarray}
\left<f_x|{\cal J}_x(P_f,P)|i_-\right> \Big|_{e_1\,{\rm
term}}&=&\frac{16 e_1N_c\,P_{-}}{{(2
\pi)\cal N}^2} \left\{
\int_{y}^1  + 
\int_{0}^{y} {\cal R}\right\} d\xi
\;\Phi_f(\xi',P_f)\Phi_i(\xi,P)\nonumber\\
&&\qquad\qquad\times
\frac{q_+\,\left[m_1^2-
\eta(1-\eta)\,Q^2 \,F_1(\eta,Q^2)\right]}
{m_{01}\,\eta\,(1-\eta)\,Q^2}
\nonumber\\
\left<f_-|{\cal J}_+(P_f,P)|i_-\right> \Big|_{e_1\,{\rm
term}}&=&\frac{16 e_1N_c\,P_{-}}{{(2
\pi)\cal N}^2} \left\{
\int_{y}^1  + 
\int_{0}^{y} {\cal R}\right\} d\xi
\;\Phi_f(\xi',P_f)\Phi_i(\xi,P)\nonumber\\
&&\qquad\qquad\times
\frac{q_+\,\left[m_1^2-
\eta(1-\eta)\,Q^2 \,F_1(\eta,Q^2)\right]}
{q_-\,\eta\,(1-\eta)\,Q^2}
\, ,\qquad\qquad
\label{currentd}
\end{eqnarray}
where the state $f_x$ has the structure $\gamma_-\gamma_x$. 
These results are used in the next subsection, and in the
discussion of DIS.

Using the definitions of the momenta shown in Fig.\
\ref{hadronicamplitude}, it is easy to see that the second term
in the transition current can be obtained from the first by
letting $1\leftrightarrow2$, as discussed in Appendix
\ref{appen:D}. The exact expression for the $x$ component of the
transition current, which is the sum of the $e_1$ and $e_2$
contributions, can therefore be written 
\begin{eqnarray}
\left<f_x|{\cal J}_x(P_f,P)|i_-\right> = {\cal
J}^{1,x}_{fi}(Q^2,y)+{\cal J}^{2,x}_{fi}(Q^2,y)
\label{exacttran}
\end{eqnarray}
where the reduced current ${\cal J}^{j,x}_{fi}(Q^2,y)$ is
\begin{eqnarray}
{\cal J}^{j,x}_{fi}(Q^2,y)=\frac{e_jm_{0j}y}{1-y}
\left\{
\int_{y}^1  + 
\int_{0}^{y} {\cal R}_j\right\} d\xi
\;\frac{\Phi_f(\xi_j',P_f)\Phi_i(\xi_j,P)}{\xi_j'\xi_j}\, 
F_{j}^{\rm eff}(\eta_j,Q^2)
\label{tranff}
\end{eqnarray}
with $\{\xi_1,\xi_1',\eta_1\}=\{\xi,\xi'\eta\}$ and $\{\xi_2,
\xi_2',\eta_2\}=\{1-\xi,1-\xi',1-\eta\}$ and
\begin{eqnarray}
F_{i}^{\rm
eff}(\eta,Q^2)=\frac{1}{m_{0i}^2}\left[m_i^2-
\eta(1-\eta)\,Q^2 \,F_i(\eta,Q^2)\right] \, . \label{qffeff}
\end{eqnarray}
Finally, the function ${\cal R}_1={\cal R}(m_1,m_2)$,
was defined in Eq.\ (\ref{Rlimit}); the substitutions that
convert the $e_1$ term into the $e_2$ term therefore give
${\cal R}_2={\cal R}(m_2,m_1)$.

\subsection{Elastic scattering and normalization of the wave
function}
\label{elasticff} 

If the scattering is elastic, $x=1$ and the momentum fraction
$y=-q_-/P_-$ must be computed from the exact expressions for $q$
and $P$ in the c.m. frame \cite{ref1}.  The result, for
elastic scattering from the state $n$, is  
\begin{equation}
y=\frac{Q^2+Q\sqrt{4M_n^2+Q^2}}{2M_n^2+Q^2+Q\sqrt{4M_n^2+Q^2}}
\label{yelas}
\end{equation}
As $Q^2\to0$, $y\to0$ and $\eta$ [defined in Eq.\ (\ref{fraccur})]
approaches $-\infty$,
$F_i(\eta,Q^2)\to0$, and the {\it exact\/} result for the
elastic curent becomes
\begin{eqnarray}
\left<n|{\cal J}_{-}(P,P)|n\right>\Big|_{e_1\,{\rm term}}
&=&\frac{16 e_1N_c\,P_{-}}{{(2
\pi)\cal N}^2}
\int_0^{1}d\xi
\;[\Phi_n(\xi,P)]^2=2\,e_1\,P_-\, {\cal F}_n(0)\, ,
\label{3eq1c}
\end{eqnarray}
where charge normalization requires that the form factor ${\cal
F}_n(Q^2)$ be unity at $Q^2=0$.  Hence
\begin{eqnarray}
{\cal N}=2\sqrt{\frac{N_c}{\pi}} \, ,
\label{3eq1d}
\end{eqnarray}
gives the normalization condition (\ref{norm}).  

Furthermore, elastic scattering requires that, as $Q^2\to0$, 
\begin{equation}
(\eta-1)\,q_-=\xi\,P_-\to\eta\,q_-
\, ,
\end{equation}
and hence
\begin{eqnarray}
\left<n|{\cal J}_{+}(P,P)|n\right>\Big|_{e_1\,{\rm term}}
&=&2e_1\,P_{+}\frac{
m_1^2}{M_n^2}\int_0^1d\xi\,\frac
{\Phi_n^2(\xi,M_0^2)}{\xi^2}
\nonumber\\
&=&2e_1\,P_+\, {\cal F}_n(0)
=2e_1\,P_+\, ,
\end{eqnarray}
which follows from the wave function identity (\ref{idet3}). 
Note that {\it the charge is properly normalized and
conserved for\/} both {\it components because we used the
correct dressed quark current\/} (\ref{current1}).  

Finally, we point out that the elastic transverse current
is zero for all $Q^2$, as required by the summetries of the
states 
\begin{eqnarray}
\left<n|{\cal J}_x(P_f,P)|n\right>
= 0\, .
\label{3eq1cc}
\end{eqnarray}
We now turn to a discussion of deep inelastic scattering.
 

\section{deep inelastic scattering}
 
\subsection{Extracting the structure functions from the DIS
cross section}
\label{DIScross}

Using the standard conventions (also defined in Ref.~\cite{ref1}),
the DIS cross section is 
\begin{eqnarray}
\frac{d^2\sigma}{d\Omega'dE'}&=&\left(\frac{2\alpha E'}
{Q^2}\right)^2
\left[W_2\cos^2\left(\frac{\theta}{2}\right)+
2W_1\sin^2\left(\frac{\theta}{2}\right)\right]\nonumber\\
&=&\sigma_M\left[W_2+2W_1\tan^2\left(\frac{\theta}{2}\right)
\right]\, ,
\end{eqnarray}
where the structure functions $W_1$ and $W_2$ are part of
the hadronic tensor, defined by
\begin{eqnarray}
\frac{W_{\mu \nu}}{4\pi M}&=&-\left( g_{\mu \nu}-
\frac{q_{\mu} q_{\nu}}{q^2} \right)W_1
+\left(P_{\mu}-q_{\mu}\frac{P \cdot q}{q^2} \right)
\left(P_{\nu}-q_{\nu}\frac{P \cdot q}{q^2} 
\right)\frac{W_2}{M^2} \nonumber\\
&=&-\left( g_{\mu \nu}-
\frac{q_{\mu} q_{\nu}}{q^2} \right)W_1
+\left(P_{\mu}+\frac{q_{\mu}}{2x} \right)
\left(P_{\nu}+\frac{q_{\nu}}{2x} 
\right)\frac{W_2}{M^2}\, ,
\label{wmn}
\end{eqnarray}
where $M=M_1$ is the target, or ground state mass.
The structure function $W_1$ is immediately extracted from
the $xx$ (or $yy$) component of the tensor  
\begin{eqnarray}
M W_1&=&\frac{W_{xx}}{4\pi}\to F_1(x)=\frac{1}{2}f(x)\, .
\label{w1}
\end{eqnarray}
Here we anticipate that $MW_1$ scales in the deep inelastic
limit to the function $F_1(x)$ of the variable $x=Q^2/(2P\cdot
q)$.  Note that $W_1$ must be identically zero if the $x$
(and $y$) components of the current are zero.

The structure function $W_2$ can be extracted from the $+$ and
$-$ components of the current using
\begin{eqnarray}
W_{-+}&=&\frac{1}{2}\left(W_{00}-W_{zz}\right)= 2\pi
M\left\{W_2\left(1+\frac{Q^2}{4x^2 M^2}
\right)-W_1 
\right\} \, .
\label{w21}
\end{eqnarray}
In the deep inelastic limit this gives    
\begin{eqnarray}
\nu W_2\to F_2(x)= 2xF_1(x) + \frac{1}{\pi}xW_{-+}\, .
\label{w2}
\end{eqnarray}
The contribution from $W_{-+}$ will be shown to go like
$\nu^{-1}$ in the following sections.  Hence, {\it if $F_1$ is
non-zero\/}, we obtain the Callan-Gross relation.  However, if
the transverse components of the current are omitted, $F_1=0$
and $\nu W_2$ does not scale.  In this case it is the quanity
$\nu^2 W_2$ that scales, and we recover the results of Einhorn
\cite{EINHORN}.

We now turn to a calculation of the structure functions.  We
calculate them first (i) in the {\it partonic picture\/}, where
all final state interations are ignored and the quarks in the
final state are assumed to be free, and then (ii) in the {\it
hadronic\/} picture, where confinement ensures that the only
possible final states are the bound $q\bar q$ states we have
already discussed.  We emphasize that the partonic picture 
can never actually occur because the confining interaction,
which also acts in the final state, can never be ignored.

\subsection{DIS in the partonic picture}
\label{DISparton}

Consider deep inelastic scattering (DIS) in the partonic 
picture, which assumes a final state composed of free quarks
with no interaction.  
%
\begin{figure}[hbt]
\leftline{
\mbox{
   \epsfxsize=4.0in\epsffile{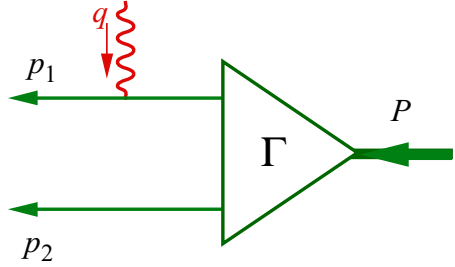}
}
}
\caption{\footnotesize\baselineskip=10pt One of two diagrams that contribute to DIS in the partonic picture.  Note the labeling of the momenta. }
\label{partonicamplitude}
\end{figure}
%
This picture is illustrated in Fig.~\ref{partonicamplitude}.
The full result is the sum of a contributions in which the
photon is absorbed by the quark with charge $e_1$, and one in
which it is absorbed by the antiquark with charge $e_2$.  In the
center of mass of the ejectiles, either particle can  go
forwards or backwards and the unpolarized cross section is the
sum over both helicities of the particles.

Begin by calculating the transverse component (chosen be be
$x$) of the current.  The outgoing quark has a mass of $m_1$ and 
momentum  $p_1$ in the center of mass frame of the ejectiles,
while the antiquark has a mass of $m_2$ and momentum of 
$p_2$ (in this section $p_2$ will be the physical momentum of
the outgoing antiquark, opposite in sign to that used in the
earlier sections of this paper). The current is the sum of two
diagrams, only one of which is shown in
Fig.~\ref{partonicamplitude}.  Since the
$q\bar q$ vertex contains the factor of $\gamma_-$, only the
factor $m+p_-\gamma_+$ in the numerator of the dressed
propagator (\ref{1eq16}) survives, but we will retain only the
factor of $p_-\gamma_+$ which dominates the result at large
$Q^2$.  The following result is obtained for the $x$ component
of the current for the disintegration of the ground state
(the subscript $n=1$ usually attached to the ground state will
be omitted in this section):
\begin{eqnarray}
J_x&=& -e_1
\left[{\bar u}(p_1)\,\gamma_x\,\gamma_+\gamma_-\,
v(p_2)\right] \frac{(p_1-q)_-\,G(p_1-q,-p_2)}{m_1^2-(p_1-q)^2
}\nonumber\\ &&+e_2\,[{\bar u}(p_1)\,\gamma_-\gamma_+ \gamma_x\,
v(p_2)] \frac{(p_2-q)_-\, G(p_1,-p_2+q) 
}
{m_2^2-(p_2-q)^2} \, .
\label{5eq1}
\end{eqnarray}
Introducing the momentum of the bound state, $P=p_1+p_2-q$,
and defining 
\begin{equation}
z=\frac{(p_1-q)_-}{P_-}=\frac{(P-p_2)_-}{P_-} \label{defza}
\end{equation}
[as in Eq.~(\ref{defz})], the first denominator can be written
\begin{eqnarray}
m_1^2-(p_1-q)^2&=& m_1^2-(P-p_2)^2=m_1^2-2
\left[\frac{M^2}{2P_-}-\frac{m_2^2}{2p_{2-}}\right](P-p_2)_-
\nonumber\\
&=& m_1^2-z\,\left(M^2-\frac{m_2^2}{1-z}\right) =
z\left[\frac{m_1^2}{z}+\frac{m_2^2}{1-z}-M^2\right]
\, .
\label{5eq1a}
\end{eqnarray}
Here we have made use of the constraints $M^2=2P_-P_+$ and
$m_2^2=2p_{2-}p_{2+}$.  A similar result follows for the
second denominator if we define a new momentum fraction $z'$   
\begin{equation}
z'=\frac{(p_2-q)_-}{P_-}=\frac{(P-p_1)_-}{P_-}\, .
\label{defzp}
\end{equation}
With these definitions we obtain 
\begin{eqnarray}
J_x&=& -e_1[{\bar u}(p_1)\,\gamma_x\,\gamma_+\gamma_-\, v(p_2)]
\;\frac{ P_-
G(z,P)}{\Delta(z,M^2)}
\nonumber\\ && +e_2 [{\bar u}(p_1)\,
\gamma_-\gamma_+ \gamma_x\, v(p_2)] \;    
 \frac{ P_- G(1-z',P) }{
\Delta(1-z', M^2)} 
\nonumber\\ \nonumber\\
&=&-\frac{e_1}{{\cal N}} \left[{\bar
u}(p_1)\,\gamma_x\,\gamma_+\gamma_-\, v(p_2) \right]
\Phi(z,P) \nonumber\\&&
+\frac{e_2}{\cal N}\left[{\bar
u}(p_1)\,
\gamma_-\gamma_+ \gamma_x\, v(p_2) \right]\Phi(1-z',P)
\, ,
\label{5eq1b}
\end{eqnarray}
where the answer has been expressed in terms of the
wave function (\ref{2eq3d}). 

The current and the DIS cross section are evaluated in the
c.m.~frame, using the kinematics defined in Appendix
\ref{appen:E}.   The unpolarized cross section is obtained by
squaring the current and summing over spins using 
\begin{eqnarray}
\sum_{\rm spins} \bar
u(p_1)\gamma_x\gamma_+\gamma_-v(p_2)\,\bar
v(p_2)\gamma_-\gamma_+\gamma_x u(p_1) &=& p_{1+}p_{2-} {\rm
trace}\left[\gamma_+\gamma_-\gamma_+\gamma_x\gamma_-
\gamma_x\gamma_+\gamma_-\right]\nonumber\\
&=&32 \,p_{1+}p_{2-} \, ,
\end{eqnarray}
and neglecting the interference term (which vanishes in the
large $Q^2$ limit).  Observing that $z\to z'\to x$ in the DIS
limit, and summing over all colors of the outgoing quarks (which
do not interfere) gives 
\begin{eqnarray}
\sum_{\rm spins} |J_x|^2&=& \frac{32N_c}{{\cal
N}^2}\left\{e_1^2\,p_{1+}^{(1)}\,p_{2-}^{(1)}\,\Phi^2(x,P) +
e_2^2\,p_{1-}^{(2)}\,p_{2+}^{(2)}\,\Phi^2(1-x,P) \right\}
\nonumber\\
&=&4\pi\,Q^2\left(\frac{1-x}{x} \right)
\left\{e_1^2\,\Phi^2(x,P) + e_2^2\,\Phi^2(1-x,P) \right\}
\, ,
\label{5eq2a}
\end{eqnarray}
where the momentum are evaluated in
Appendix \ref{appen:E}. The hadronic tensor defined in
Eq.~(\ref{wmn}) is related to the square of the currents by 
\begin{eqnarray}
\frac{W_{\mu \nu}}{4\pi M}&=&\int \frac{dp_1dp_2}{(2
\pi)^2\,4p_{10}p_{20}} (2 \pi)^2 \,\delta^{2}(P_f-P-q)
\sum_{\rm spins} J_{\mu}J_{\nu}\, ,
\label{4eq5a}
\end{eqnarray}
The $W_{xx}$ component of the tensor is \cite{ref1} is therefore
\begin{eqnarray}
W_{xx}&=&4\pi\,M\,W_1=\frac{1}
{4\, |p_z|(p_{10}+p_{20})}\sum_{\rm spins} |J_x|^2 
\nonumber\\
&=&2\pi\,
\left\{e_1^2\,\Phi^2(x,P) + e_2^2\,\Phi^2(1-x,P) \right\}
\, .
\label{5eq2b}
\end{eqnarray}
This gives the familiar parton model
result for the structure function $W_1$   
\begin{eqnarray}
M\,W_1= F_1(x)=\frac{1}{2}\,
\left\{e_1^2\,\Phi^2(x,P) + e_2^2\,\Phi^2(1-x,P) \right\}
=\frac{1}{2}f(x)
\, .\label{5eq2c}
\end{eqnarray}

The structure function $W_2$, given in Eq.~(\ref{w2}), is then 
\begin{eqnarray}
\nu\,W_2=2MxW_1=2xF_1(x)=F_2(x)=xf(x)\, ,
\label{5eq2e}
\end{eqnarray} 
where we anticipate the result
$W_{-+}\to0$ as $Q^2\to\infty$. This is the Callan-Gross
relation.

Now look at the plus and minus components of the current.  If
the Dirac space were restricted to two dimensions, only these
components of the current would exist.  The minus
component can be obtained immediately from (\ref{5eq1b}) by
replacing
$\gamma_x$ with $\gamma_-$, giving
\begin{eqnarray}
J_-&=&-\left[{\bar u}(p_1)\,
\gamma_-\, v(p_2) \right]\,\frac{2}{\cal N}\,\left\{e_1
\Phi(z,P)-e_2 \Phi(1-z',P)\right\}
\, ,
\label{5eq1d}
\end{eqnarray}
The plus component could include the two contributions from the
quark current, Eq.~(\ref{current1}), but in the partonic
picture the outgoing quark is not dressed, so the second part
of the current proportional to $g^2$ should be ignored.  However,
in order to study the contribution of this term we will retain
it for now. The leading contribution from both terms in the
$J_+$ current is 
\begin{eqnarray}
&J_+&=-\frac{e_1}{\cal N}\left\{m_{01}
\frac{\Phi(z,P)}{p_{1-}^{(1)}-q_-} \,\left[{\bar
u}(p_1)\,
\gamma_+\,\gamma_-\, v(p_2)
\right]+ \frac{g^2}{\pi}\frac{\Phi(z,P)}
{p_{1-}^{(1)}(p_{1-}^{(1)}-q_-)}\,\left[{\bar
u}(p_1)\,\gamma_-\, v(p_2)\right]
 \right\}\nonumber\\ 
&& +\frac{e_2}{\cal N}\left\{m_{02}
\frac{\Phi(1-z',P)}{p_{2-}^{(2)}-q_-}\,\left[{\bar
u}(p_1)\,
\gamma_-\,\gamma_+\, v(p_2) \right] + \frac{g^2}{\pi}
\frac{\Phi(z,P)} {p_{2-}^{(2)}(p_{2-}^{(2)}-q_-)}\,
\left[{\bar u}(p_1)\,\gamma_-\, v(p_2)\right] \right\}
 .\qquad\quad
\label{5eq1e}
\end{eqnarray}
The $W_{-+}$ component of the hadron tensor is then
\begin{eqnarray}
W_{-+} = \frac{1}
{4\, |p_z|(p_{10}+p_{20})}\sum_{\rm spins} J_-J_+ =
\frac{x}{2Q^2(1-x)}\sum_{\rm spins} J_-J_+\, . 
\label{5eq1f}
\end{eqnarray}
Using (\ref{5eq1d}) and (\ref{5eq1e}), and summing over colors,
the $e_1^2$ contribution to the current sum becomes, at large
$Q^2$,
\begin{eqnarray}
\sum_{\rm spins} J_-J_+\Big|_{e_1\,{\rm term}}
&=&\frac{32\,e_1^2\,N_c}{{\cal N}^2}\,
\Phi^2(x,P)\,\left[\frac{p_{2-}^{(1)}}
{p_{1-}^{(1)}-q_-}\right]\left(m_{01}^2-\frac{g^2}{\pi}\right)
\nonumber\\
&=&8\pi\,e_1^2\,m_{1}^2
\left(\frac{1-x}{x}\right)
\;\Phi^2(x,P)\, .
\label{5eq1g}
\end{eqnarray}
Hence
\begin{eqnarray}
W_{-+}\big|_{e_1\,{\rm term}}
&=&\frac{4\pi\,e_1^2}{Q^2}\,m_{1}^2
\;\Phi^2(x,P)=4\pi\frac{m_1^2}{Q^2}f(x)\big|_{e_1\,{\rm
term}}
\, ,\label{5eq1h}
\end{eqnarray}
and this term is subleading at high
$Q^2$.  Note that the effect of neglecting the term
proportional to $g^2$ in the $+$ component of the quark
current (as we are instructed to do in the partonic picture) is
to replace $m_1^2$ by the bare quark mass $m_{01}^2$. If
$W_1=0$, then $\nu^2W_2$ scales as  
\begin{eqnarray}
\nu^2W_2\big|_{e_1\,{\rm term}}\to \frac{Q^2}{2\pi
M}W_{-+}\big|_{e_1\,{\rm term}}
&=&2\frac{m_{01}^2}{M}f(x)\big|_{e_1\,{\rm
term}}
\, .  \label{einhorn}
\end{eqnarray}

Up to a factor, this is the result originally obtained by
Einhorn \cite{EINHORN}.  In that paper, the Dirac space was
restricted to 2 dimensions, $W_1$ was necessarily zero, and
$\nu^2 W_2$ scaled as (\ref{einhorn}).  This distribution
amplitude does not have the correct physical interpretation. 
In particular, it predicts that the DIS cross section depends
on the square of bare quark mass, a clearly unphysical result. 
Only by considering the full four dimensional Dirac space are
we able to obtain scaling for $\nu W_2$, and the familiar
physical result (\ref{5eq2c}) for $f(x)$.  
 
Another consequence of the fact that $W_{-+}$ is
subleading is that we can avoid coming to grips with the fact
that the current components $J_-$ and $J_+$, as
defined above, do not conserve current.  Since these
components do not contribute to the final result, we need
not discuss how they can be redefined in order to conserve
current.

\subsection{DIS in the hadronic picture}

\subsubsection{Exact results for inelastic scattering}

The {\it exact\/} hadronic tensor (\ref{wmn}) for
electroproduction of a {\it single\/} final state $f$
with mass $M_n$ (denoted by $n$) from the ground state $i$
with mass $M_1\equiv M$ (denoted by 0) is 
\begin{eqnarray}
W_{\mu \nu}\bigl|_f&=&\int \frac{dP_{fz}}{(2 \pi)2 P_{f0} }
(2 \pi)^2\,\delta^{2}(P_f-P-q)\sum_{\rm spins}
\left<0|{\cal J}_{\mu}|n\right>\left<n|{\cal J}_{\nu}|0\right>
\nonumber\\
&=&\delta(M_n-P_0-q_0)\;\frac{\pi}{M_n} \sum_{\rm spins}
\left<0|{\cal J}_{\mu}|n\right>\left<n|{\cal J}_{\nu}|0\right> \,
.
\label{tensor2}
\end{eqnarray}
This is a delta function.  In any
physical measurement, the detectors will accept a
finite range of values of the final electron energy, $E'$,
requiring that this theoretical cross section be averaged over
$E'$.  This averaging process will be discussed in 
subsection \ref{sec:duality} below, where the deta function will
be written as a delta function in the Bjorken variable $x$, which
spikes at values of $x_n$ corresponding to the excitation of
the final state $n$.  In the c.m. system, this leads to the
following {\it exact\/} transformation
\begin{eqnarray}
\delta(M_n-P_0-q_0)=\delta(x-x_n)\frac{2x_n^2M_n}{Q^2} 
\label{xdef}
\end{eqnarray}
with $x_n$ the value of $x$ at the bound state mass
$M_n$, given in Eq.~(\ref{xf}) below. 
 
In the DIS limit, the transverse currents dominate the
scattering.  The exact result for the transverse tensor,
$W_{xx}$, that describes the scattering from the ground state
with $\gamma_-$ structure, to a {\it single\/} transverse final
state $n$ with $\gamma_-\gamma_x$ structure, is  
\begin{eqnarray}
W_{xx}\bigl|_n=\delta(x-x_n)\;\frac{2\pi x_n^2}{Q^2}
\biggl\{
{\cal J}^{1,x}_{n}(Q^2,y) +{\cal J}^{2,x}_{n}(Q^2,y)
\biggr\}^2\, , 
\label{tensorxx}
\end{eqnarray}
where the reduced transverse current was defined in Eq.\
(\ref{tranff}). (Here the $fi$ subscript is replaced by $n$, and
denotes a transition between the ground and $n$th
state.)  Next we observe that the value of
$x$ at which the final state is excited is   
\begin{equation}
x_n=\frac{Q^2}{M_n^2-M^2+Q^2}\, ,\label{xf}
\end{equation}
and that the spacing between neighboring levels is
\begin{eqnarray}
\Delta_{x_n}&=&x_n-x_{n+1}=
\frac{\delta M_{n}^2}{Q^2} \;x_n\,x_{n+1}
\, ,
\label{dxf}
\end{eqnarray}
with $\delta M_n^2=M_{n+1}^2-M_n^2$. 
Hence the {\it exact\/} result for the total inelastic tensor
$W_{xx}$ is 
\begin{eqnarray}
W_{xx}=\sum_n\delta(x-x_n)\;\Delta_{x_n} \left(\frac{2\pi}
{\delta M_{n}^2}\right)\left(\frac{x_n}{x_{n+1}}\right)
\biggl\{
{\cal J}^{1,x}_{n}(Q^2,y) +{\cal J}^{2,x}_{n}(Q^2,y)
\biggr\}^2\, , 
\label{tensorxx2}
\end{eqnarray}
where the sum is over all possible final states $n$.

\subsubsection{Hadronic result in the DIS limit}

The DIS limit is defined by the requirement that $Q^2\to\infty$
with $x$ held constant.  In this limit both $Q^2$ and $M_n^2$
become large, with
\begin{eqnarray}
x_n\to\frac{Q^2}{M_n^2+Q^2}\, ,\qquad({\rm as}\;
Q^2 \;{\rm and}\; M_n^2\to\infty)\, .
\label{xf2}
\end{eqnarray}
Study of the DIS limit therefore requires estimating the
transition form factors for large final state number $n$, but for
all $x$ (since $x_n$ depends on $Q^2$).  In the DIS limit, $y\to
x$ [easily seen from the expansions (\ref{5eq5})], so
$y$ will be replaced by $x$ in the following discussion.

When $n$ is large the ``tails'' of the wave functions 
are very small, and the exact transition current (\ref{tranff})
can be approximated  by neglecting the contribution from the
region $\xi\in[0,x]$ (where $\xi'<0$), giving
\begin{eqnarray}
{\cal J}^{1,x}_{n}(Q^2,x)&\simeq&\frac{e_1\,m_{01}\,x}{1-x}
\int_{y}^1  d\xi
\;\frac{\Phi_n(\xi',P_f)\Phi(\xi,P)}{\xi'\xi}\, 
F_{1}^{\rm eff}(\eta,Q^2)\nonumber\\
&=&e_1\,m_{01}\,x
\int_{0}^1  d\xi'
\;\frac{\Phi_n(\xi',P_f)\Phi([\xi'(1-x)+x],P)}
{\xi'[\xi'(1-x)+x]}\,  F_{1}^{\rm
eff}(\eta,Q^2)
\, , \qquad\label{translim}
\end{eqnarray}
where  we continue to suppress the label $n=1$ on the
ground (initial) state wave function, and
\begin{eqnarray}
\eta=-\xi'\,\left(\frac{1-x}{x}\right) <0
\end{eqnarray}
in the region of integration.  When $\eta<0$ and $Q^2\to\infty$, 
the relation (\ref{ff3})  may be used to approximate the quark
form factor by     
\begin{eqnarray}
F_i(\eta,Q^2)\to -\frac{g^2}{\pi\, Q^2\eta\,(1-\eta)}\, ,
\label{fflimit}
\end{eqnarray}
giving 
\begin{eqnarray}
F_1^{\rm eff}(\eta,Q^2)\to 1\,  .
\end{eqnarray}

Equation\ (\ref{translim}) is now further reduced by
expanding $\Phi(\xi,P)/\xi$ around $\xi=x$
\begin{eqnarray}
\frac{\Phi(\xi,P)}{\xi}=\frac{\Phi(x,P)}{x}
+\delta\,\frac{d}{dx}
\left(\frac{\Phi(x,P)}{x}\right)
+\frac{1}{2}\,\delta^2\,\frac{d^2}{dx^2}
\left(\frac{\Phi(x,P)}{x}\right)+\cdots\, .
\end{eqnarray}
where the higher order terms are proportional to powers of
$\delta$ 
\begin{eqnarray}
\delta\equiv\xi'(1-x) \, .
\label{delta}
\end{eqnarray}
This replacement gives the following series for ${\cal J}$
\begin{eqnarray}
{\cal J}^{1,x}_{n}(Q^2,x)&=&e_1\,m_{01}\,\Phi(x,P)
\int_{0}^1  d\xi'
\;\frac{\Phi_n(\xi',P_f)}
{\xi'} \nonumber\\ &&+e_1\,m_{01} \sum_{m=1}^{m=\infty} d_n^m\,
\frac{(1-x)^m}{m!}\left(\frac{d}{dx}\right)^m 
\left(\frac{\Phi(x,P)}{x}\right)
\label{translimseries}
\end{eqnarray}
where
\begin{eqnarray}
d_n^m=\int_0^1d\xi'\,\xi'^{m-1}\, \Phi_n(\xi',P_f)\, .
\label{dmn}
\end{eqnarray}
Using the identity (\ref{idet4}) and the other estimates given in
Appendix \ref{appen:A2} it follows that 
\begin{eqnarray}
\lim_{n\to\infty} \frac{d_n^m}{C_n} \to \frac{{\rm const}}{n}
\to0  \, .
\label{dmn2}
\end{eqnarray}
Hence the large $n$ approximation to (\ref{translimseries}) is
\begin{eqnarray}
{\cal J}^{1,x}_{n}(Q^2,x) &\to& e_1\,m_{01} \,\Phi(x,P)
\int_{0}^1  d\xi'
\;\frac{\Phi_n(\xi',P_f)}
{\xi'}\nonumber\\
&=&  e_1\,C_n \,\Phi(x,P)
\, . \label{translim2}
\end{eqnarray}

The plus component of the transition current is smaller than the
transverse current by a factor of $m_{01}/q_-$ [compare the two
results in Eq.~(\ref{currentd})], and hence is obtained
immediately from the result (\ref{translim2}).  Using
Eq.~(\ref{5eq5}) for
$q_-$, and Eq.~(\ref{xf2}) to replace $\sqrt{(1-x_n)/x_n}$,
gives  
\begin{eqnarray}
{\cal J}^{1,+}_{n}(Q^2,x) \simeq -
\sqrt{2}\,e_1\,C_n\,\frac{m_{01}\,M_n}{Q^2}\,
\Phi(x,P) \, .
\label{translimplus}
\end{eqnarray}
The minus component can also be obtained from  
Eq.~(\ref{translim2}). Reviewing the derivation, we see that
the minus component, in the DIS limit, is obtained from the
transverse component by replacing  
\begin{eqnarray}  
r\equiv\frac{m_{01}}{2\eta\,(1-\eta)\,q_-}=\frac{m_{01}\,y}
{\sqrt{2}\,\xi'\xi\, M_n}\to 1
\end{eqnarray}
Hence, multiplying (\ref{translim2}) by $r^{-1}$ 
\begin{eqnarray}
{\cal J}^{1,-}_{n}(Q^2,x) &\simeq&\sqrt{2}\,e_1\,\Phi(x,P)
\left\{M_n\int_0^1d\xi^*
\;\Phi_f(\xi^*,P_f) \right\}\nonumber\\
&\simeq&\sqrt{2}\,e_1\,
C_n\left(\frac{m_{01}+(-1)^{n-1}
m_{02}}{M_n}\right)\,\Phi(x,P)\, ,
\label{translimminus}
\end{eqnarray}
where the $\xi'$ integral was evaluated using identity
(\ref{B10}).  

The transition currents can be shown to be gauge invariant by
useing the Ward-Takahashi identity on  the quark-photon vertex,
and reducing the result using the Bethe-Salpeter equation.
 
These currents will now be used to determine the structure
functions $W_1$ and $W_2$ in the hadron picture.

\subsubsection{The DIS cross section from the transition form
factors}

The cross section in the DIS limit can now be obtained
from the exact formula  (\ref{tensorxx2}).  As long as $x$ is not
too close to unity each term in the sum must correspond to some
state with large $n$ (the states with small $n$ ``pile up'' near
$x=1$ where their total contribution is very small), and the
approximate result for the transverse current,
(\ref{translim2}), may be used.  Furthermore, for large $n$,
$\delta M_n^2$ may be approximated by using (\ref{B11}), and
$x_{n+1}\simeq x_n$.  Hence the final hadronic result [denoted
$\tilde f$ to distinguish it from the partonic result $f$
of Eq.\ (\ref{5eq2c})] in the DIS limit is
\begin{eqnarray}
\tilde f(x)=2\tilde F_1(x)=\frac{W_{xx}}{2\pi}=\sum_n\delta(x-x_n)
\;\Delta_{x_n}
\biggl\{e_1\,\Phi(x,P) + (-1)^ne_2\Phi(1-x,P)
\biggr\}^2\, .\qquad 
\label{f1}
\end{eqnarray}
Note that $\tilde F_1(x)$ is a series of separate spikes, zero
for all $x$ except at particular values $x_n$, where it is
infinite.  The partonic function, $F_1(x)$, is a
smooth function of $x$.  To compare $\tilde F_1(x)$ to 
$F_1(x)$, we must average over $x$.  This will lead to the
concept of duality and will be discussed in the next subsection
below.     

Now look at the $W_{-+}$ component of the tensor.  Replacing the
transverse components of the current in Eq.~(\ref{tensorxx2})
with plus and minus components, and taking the DIS limit, 
gives 
\begin{eqnarray}
W_{-+}&=&\frac{2\pi}{C_\infty^2}\sum_n\delta\left(x-x_n\right)
\Delta_{x_n}\nonumber\\
&&\times\left\{{\cal J}^{1,-}_{n}(Q^2,x) + {\cal
J}^{2,-}_{n}(Q^2,x)\right\}\left\{{\cal J}^{1,+}_{n}(Q^2,x) +
{\cal J}^{2,+}_{n}(Q^2,x)\right\}^*\, . 
\label{wpm}
\end{eqnarray}
Substituting Eqs.~(\ref{translimplus}) and (\ref{translimminus}),
and droping the term proportional to $(-1)^{n-1}$ (which averages
to zero) gives 
\begin{eqnarray}
W_{-+}\big|_{e_1\,{\rm
term}}&=&\frac{4\pi\,m_{01}^2}{Q^2}\sum_n 
\delta\left(x-x_n\right)\,
\Delta_{x_n}\,e_1^2
\,\Phi^2(x,P) \, .
\label{wpm2}
\end{eqnarray}
This is a nonleading term and can be neglected in the DIS
limit.  Hence the hadronic picture also gives the 
Callan-Gross relation, and it is sufficient to compare the
functions $F_1$ and $\tilde F_1$ only. 

If $W_1$ were zero, Eq.~(\ref{wpm2}) would give the following
scaling relation for $\nu^2W_2$
\begin{eqnarray}
\nu^2W_2\big|_{e_1\,{\rm term}}\to \frac{Q^2}{2\pi
M}W_{-+}\big|_{e_1\,{\rm term}}
&=&2\frac{m_{01}^2}{M}\,\tilde f(x)\big|_{e_1\,{\rm
term}}
\, .\qquad\qquad\label{einhorn2}
\end{eqnarray}
This is to be compared with (\ref{einhorn}).  Hence, the
duality of $W_1$ and the duality of the nonleading terms in
$\nu^2W_2$ depends on comparison of the same functions, $f(x)$
and $\tilde f(x)$.

We discuss this comparison now.


\subsection{Duality and its implications}
\label{sec:duality}

In any physical measurement, the detectors which define the
final state will accept a range of final electron energies
$\delta E'$.  For fixed $Q^2$, this can be converted into the
acceptance of a range of values of $x$ centered at $x_i$ with
width $\delta x$ so that $x$ varies over the interval bounded
by $x_{i\pm}=x_i\pm\delta x/2$.  Then the {\it experimentally
measured\/} hadronic structure function in the DIS limit can be
computed from (\ref{f1})  
\begin{eqnarray}
\left<\tilde f\right>_{x_i} &\equiv& \frac{1}{\delta x} 
\int_{x_{i-}}^{x_{i+}} dx \tilde f(x) \nonumber\\
&=&
\frac{1}{\delta x} \sum_{n\,\in\, x_i}
\Delta_{x_n}\biggl|e_1 \Phi(x_n,P) +(-1)^{n-1} e_2\,
\Phi(1-x_n\, ,P)
\biggr|^2 \label{f1hadronb}\, .
\end{eqnarray}
Note that this is the sum of smooth terms proportional to $e_1^2$
and $e_2^2$, and a rapidly oscillating interference term
proportional to $e_1e_2$.  In the DIS limit, the separation
$\Delta_x$ between states approaches zero, and therefore a
large number of states are necessarily included in any interval
$\delta x$.  The interference term, which changes sign as
$(-1)^{n-1}$, therefore averages to zero.  This cancellation of the
interference term was pointed out originally by Einhorn
\cite{EINHORN}, and empahsized recently by Close and Isgur in the
context of the nonrelativistic quark model
\cite{CI}. 
 
The smooth terms can be approximated by their value at the center
of the interval.  Using the fact that the number of states in the
interval times the $x$ spacing between them must necessarily
equal the width $\delta x$, so that
\begin{eqnarray}
\sum_{n\,\in\, x_i} \Delta_{x_n} = \delta x\, . \label{space}
\end{eqnarray}
we can reduce (\ref{f1hadronb}) to
\begin{eqnarray}
\left<\tilde f\right>_{x_i}&\simeq&\biggl\{e^2_1 \Phi^2(x_i,P) +
e^2_2 \Phi^2(1-x_i,P) \biggr\} 
\frac{1}{\delta x} \sum_{n\,\in\, x_i} \Delta_{x_n}
\nonumber\\
&&+2e_1e_2\,\frac{1}{\delta x} \sum_{n\,\in\, x_i} (-1)^{n-1}
\Delta_{x_n} \Phi(x_n,P) \Phi(1-x_n,P) \nonumber\\
&\to& e^2_1 \Phi^2(x_i,P) + e^2_2 \Phi^2(1-x_i,P)=f(x_i)\, ,
\label{f1hadronc}
\end{eqnarray}
where $f(x)$ was defined in Eq.\ (\ref{5eq2c}). 

\begin{figure}
\centerline{
\mbox{
   \epsfxsize=5.0in\epsffile{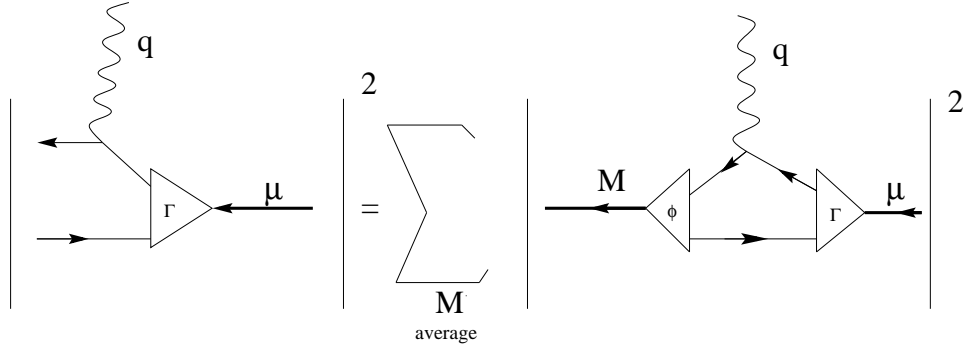}
}}
\caption{\footnotesize\baselineskip=10pt Quark-hadron duality}
\label{hadronic}
\end{figure}


This result should be compared with the partonic result, averaged
in the same way.  Since the partonic result is smooth (and the
interference term already neglected) the partonic average is
trival 
\begin{eqnarray}
\left<f^{\phantom{0}}\!\!\right>_{x_i} \equiv \frac{1}{\delta x} 
\int_{x_{i-}}^{x_{i+}} dx f(x) \simeq f(x_i) \,
\frac{1}{\delta x}  \int_{x_{i-}}^{x_{i+}} dx
=f(x_i)\, .
\label{f1parton}
\end{eqnarray}
Hence we see that  
\begin{eqnarray}
\left<\tilde f\right>_{x_i} =
\left<f^{\phantom{0}}\!\!\right>_{x_i}
\, ,
\label{f1hadron}
\end{eqnarray}
proving duality. 

Note that a similar identity holds for the $n$th
moments of $f$.  In the high $Q^2$ limit when the final states are
numerous and close enough together so that
$\Delta_{x_n}\to dx\to0$,  the $m$th moment of the
hadronic structure function is 
\begin{eqnarray}
\left<x^m\right>\equiv\int_0^1dx\;x^m\tilde f(x)
&=&\sum_n\,\Delta_{x_n}\;
x_n^{m}\;\left[e^2_1\,\Phi^2(x_n,P) 
+e_2^2\,\Phi^2(1-x_n\, ,P) \right]
\nonumber\\ &\to& \int_0^1 dx\;
x^{m}\;\left[e^2_1\,\Phi^2(x,P) +e^2_2\,\Phi^2(1-x\, ,P)
\right]\nonumber\\ &=&\int_0^1dx\;x^m f(x)
\, .
\label{f1a}
\end{eqnarray}

Quark-hadron duality means that properly averaged hadronic
observables in a certain kinematic regime (high $Q^2$) can 
be described by perturbative QCD, as schematically 
represented in Fig.\ \ref{hadronic} and demonstrated in Eq.\
(\ref{f1hadron}).  Recent data from Jefferson Laboratory exhibit
duality, and this has focused attention on this old subject.
Recent studies show that the
weak decays within the 't Hooft model exhibit duality
\cite{gl,lu,l}, and in the framework of DIS, Close and Isgur
\cite{CI} studied its emergence in the context of realistic
nonrelativistic quark models. A recent paper by Isgur,
Jeschonnek,  Melnitchouk and Van Orden
\cite{IJMVO} studied duality using a relativistic
Klein-Gordon equation with a confining interaction.  This work
is complementary to these other studies.

While we have proved that duality emerges in the DIS limit, and
have corrected the pathologies of earlier work, we have not
yet studied how duality emerges as a function of $Q^2$. A future paper will present  numerical studies of the onset of duality for
rQCD \cite{BGII}.


\section{Conclusions}

We have shown that a very satisfactory description of deep inelastic
scattering can be obtained from large $N_c$ QCD in 1+1 dimensions, {\it provided\/} the quark spin degrees of freedom are extended to the full 1+3 dimensions (along with the leptons and the electromagnetic field).  We refer to this new model as reduced QCD, and suggest that rQCD may be used to provide insight into the behavior of many other physical processes.  In the applications to DIS discussed in this paper, we are able to confirm the Callan-Gross relation, and that duality emerges in the deep
inelastic limit.  

Our theoretical discussion teaches several lessons. 
Electromagnetic gauge invariance and duality emerge only if
the quark current is fully and consistently dressed.  The way in which a
quark form factor emerges from this dynamical model mimics the
phenomenlogical approach of Gross and Riska \cite{RG}, showing
that their prescription has a dynamical justification.  Our treatment of confinement
removes all singularities, giving a finite mass for the
dressed quark.  This mass is gauge dependent, showing that
it is not a physical quantity, which can be taken as a
demonstration of confinement.  But the fact that we have
finte dressed quark masses in the presence of confinement
provides justification for the constituent quark model.  The
simultaneous presence of confinement and finite quark mass is
possible only if the ``on-shell'' quark scattering amplitude is
identicallyt zero, and we have shown that this is indeed true,
as modeled previously in Ref.\ \cite{SG}. 

The treatment of reduced QCD presented here lays a
foundation for much further study.

\section{ACKNOWLEDGMENTS}
This work has been supported in part by 
the DOE through grant No.\  DE-FG05-88ER40435.
The Southeastrn Universities Research Association 
(SURA) operates the Thomas Jefferson National Accelerator Facility under 
DOE contract DE-ACO5-84er40150.
One of us (Z. B.) has been supported by a Predoctoral 
Fellowship at the College of William \& Mary and by the Funda{\c c}{\~ a}o para a Ci{\^ e}ncia e a Tecnologia under the
post doctoral grant SFRH/BPD/5661/2001 the project grant CFIF-Plurianual.

\appendix
\section{Reduction of the two quark interaction}
\label{appen:A}

Integrals over the variable $k_+$ like the one encountered in
Eq.~(\ref{BSE2}) appear several places in this paper.  Here we
evaluate it in full generality. 

Consider the integral
\begin{eqnarray}
I_2(p_-,r)=4i\int 
\frac{d^2k}{(2 \pi)^2} 
\frac{V(p,k)\,k_-(k_--r_-)F(k_-,r)}
{\left[m_1^2-k^2 -i\epsilon\right] \left[m_2^2-(k-r)^2
-i\epsilon\right]}\, . \label{int1}
\end{eqnarray}
Substituting for $V$ gives
\begin{eqnarray}
I_2(p_-,r)&=&\frac{ig^2}{ \pi^2} \int 
\frac{d^2k}{(p_--k_-)^2} \left\{
\frac{F(k_-,r)}{d_2(k_-,k_+,r)}-\frac{F(p_-,r)}
{d_2(p_-,k_+,r)}
\right\}\, ,
\end{eqnarray}
where the denominator $d_2$ is 
\begin{eqnarray}
d_2(k_-,k_+,r)&=&\left[\frac{m_1^2}{k_-}-2k_+
-i\,\frac{\epsilon}{k_-}\right]
\left[\frac{m_2^2}{(k-r)_-}-2(k-r)_+
-i\,\frac{\epsilon}{(k-r)_-}\right]\nonumber\\
&=&\left[\frac{m_1^2}{y\,r_-}-2k_+
-i\,\frac{\epsilon}{y\,r_-}\right]
\left[-\frac{m_2^2}{(1-y)\,r_-}-2(k-r)_+
+i\,\frac{\epsilon}{(1-y)\,r_-}\right]\, ,\qquad \label{d2}
\end{eqnarray}
and we have introduced the momentum fractions (\ref{defz}). 
This shows that the denominator has only two poles in $k_+$,
and that they will both be in the same half of the complex
plane (giving zero for the integral) unless 
\begin{eqnarray}
0\le y\le1\, .  \label{r1}
\end{eqnarray}
Doing the integral over $k_+$, and expressing the answer
in terms of the momentum fractions $y$ and $z$ gives
\begin{eqnarray}
I_2(z\,r_-,r)\to I_2(z,r)&=&\frac{g^2}{\pi}  
\int_0^{1}\frac{dy}{(z-y)^2}\left\{
\frac{F(y\,r_-,r)}{\Delta(y,r^2)}-\theta(z)\theta(1-z)\,
\frac{F(z\,r_-,r)}{\Delta(z,r^2)}\right\}
  \nonumber\\
&&-\theta(z)\theta(1-z)\,
\frac{g^2\,F(z\,r_-,r)}{\pi\,\Delta(z,r^2)} 
\left\{\int_{-\infty}^0 +
\int_{1}^\infty\right\}
\,\frac{dy}{(z-y)^2} \, ,
\label{2eqaa}
\end{eqnarray}
where the new denominator $\Delta(y,r^2)$ is 
\begin{equation}
\Delta(y,r^2)=\left[{\displaystyle\frac{m_1^2}{y}+
\frac{m_2^2}{(1-y)}-r^2}\right] \, .\label{d1}
\end{equation}
[Note that this result holds for {\it both\/} signs of $r_-$. 
This is because the sign of the countor integral, which depends
on the sign of $r_-$, is cancelled by the sign of the integral
over $k_-=y\,r_-$, which must be changed from
$\int_1^0\to-\int_0^1$ when $r_-<0$.]  It is important to note
that the integral over
$y$ is not restricted in the term in (\ref{2eqaa}) that is
proportional to $F(z\,r_-,r)$ because for that term the
restriction (\ref{r1}) applies to $z$ and not
$y$. This term not only regulates the singularity in
the first term (which arises only if $0\le z\le 1$), but
also contibutes an additional contribution.  Furthermore, if
$z$ lies outside of the interval $0\le z\le 1$, the first
term is non singular (except at the end points of the interval)
and is nonzero!  Hence, in the region $0\le z\le 1$ the
integral (\ref{2eqaa}) becomes
\begin{eqnarray}
I_2(z,r)&=&{\displaystyle \frac{g^2}{\pi}  
\int_0^{1}\frac{dy}{(z-y)^2}\left\{
\frac{F(y\,r_-,r)}{\Delta(y,r^2)}-
\frac{F(z\,r_-,r)}{\Delta(z,r^2)}\right\}
-\frac{g^2\,F(z\,r_-,r)}{\pi\,z(1-z)\,
\Delta(z,r^2)}}\nonumber\\ &=&
\int_0^1dy\;{\rm V}(z,y)\,\frac{F(y\,r_-,r)}{\Delta(y,r^2)}
-\frac{g^2/\pi}{z(1-z)}\;\frac{F(z\,r_-,r)}{\Delta(z,r^2)}
\qquad\left({\rm if}\quad z\in [0,1]\right)\, ,\qquad\quad
\label{2eqab}
\end{eqnarray}
and outside of this region it is
\begin{eqnarray}
I_2(z,r)={\displaystyle \frac{g^2}{\pi}  
\int_0^{1}\frac{dy}{(z-y)^2}\,
\frac{F(y\,r_-,r)}{\Delta(y,r^2)}}=\int_0^1dy\;{\rm
V}_0(z,y)\;\frac{F(y\,r_-,r_-)}{\Delta(y,r^2)}
\quad\left({\rm if}\quad z\notin [0,1]\right)\, ,\qquad
\label{2eqac}
\end{eqnarray}
where we have defined  
\begin{eqnarray}
\frac{(r_-)^2 }{\pi}\,V(p,k)&\to& \cases{{\rm V}(z,y)\equiv
{\displaystyle\frac{g^2}{\pi}
\left\{\frac{1}{(z-y)^2}-\delta(z-y)\int_0^1dy'
\frac{1}{(z-y')^2}\right\} }&if $z\in[0,1]$\cr
{\rm V}_0(z,y)\equiv
{\displaystyle\frac{g^2}{\pi}
\frac{1}{(z-y)^2}} &if $z\notin[0,1]$} \, .\qquad
\label{eq251a}
\end{eqnarray}
The multiplication by $r_-^2$ in (\ref{eq251a}) is needed
to scale the momenta from $k_-\to y$, etc.  

We emphasize that the identities (\ref{2eqab}) and
(\ref{2eqac}) hold for both $r_->0$ and $r_-<0$.  The first
case is needed for the bound state equation and the second for
electron scattering.


\section{identities involving bound state wave functions}
\label{appen:A2}

In this appendix we derive a number of identities that
hold when $\Phi_n(0,r)=\Phi_n(1,r)=0$ (i.e. in all cases but the
chiral limit). Equation (\ref{2eqee}) leads immediately to the
following identity
\begin{eqnarray}
\int_0^1dz\,\left(\frac{m^2_1}
{z}+\frac{m^2_2}{1-z}\right)\,\Phi'_n(z,r)\,\Phi_n(z,r)=0
\, . \label{idet1}
\end{eqnarray}
Integrating by parts gives 
\begin{eqnarray}
m_1^2\int_0^1dz\,\frac{\Phi^2_n(z,r)}
{z^2}=m_2^2\int_0^1dz\,\frac{\Phi^2_n(z,r)}{(1-z)^2}
\, . \label{idet2}
\end{eqnarray}
An additional identity is derived by multiplying (\ref{2eqee})
by $2\,(1-z)\,\Phi_n'(z,r)$ and integrating.  First observe that
\begin{eqnarray}
&&2\int_0^1dz\,(1-z)\,\Phi_n'(z,r)\; 
\dashint_0^1dy\frac{\Phi'_n(y,r)}{(z-y)} \nonumber\\
&&\qquad\qquad=\int_0^1\int_0^1dz\,dy\;
\Phi_n'(z,r)\,\Phi_n'(y,r)\;\frac{y-z}{z-y}=0
\, . \label{potidet}
\end{eqnarray}
Hence, integrating by parts and using the normalization
condition (\ref{norm}) for the wave function, we obtain
\begin{eqnarray}
M_n^2=
m_1^2\int_0^1dz\,\frac{\Phi_n^2(z,r)}{z^2}\, .
\label{idet3}
\end{eqnarray}
Similarly we can prove that
\begin{eqnarray}
M_n^2=
m_2^2\int_0^1dz\,\frac{\Phi_n^2(z,r)}{(1-z)^2}\, .
\label{idet3b}
\end{eqnarray}

Additional identities are needed for the study of
duality.  First, integrate Eq.\ (\ref{phiin})
over $z$ to obtain
\begin{eqnarray}
M_n^2\int_0^1dz\,\Phi_n(z,r)=\int_0^1dz\left(
\frac{m_{01}^2}{z} +
\frac{m_{02}^2}{1-z}\right)\Phi_n(z,r)\, . \label{B6}
\end{eqnarray}
Next, the commutator of the operator
\begin{eqnarray}
{\cal K}\Phi_n(x,r)\equiv \dashint_0^1dy
\,\frac{\Phi_n(y,r)}{y-x} 
\end{eqnarray}
with the Hamiltonian defined in (\ref{phiin}) is
\begin{eqnarray}
[H,{\cal K}]\,\Phi_n(x,r)=\int_0^1dy \left[
\frac{m^2_{01}}{xy}
-\frac{m^2_{02}}{(1-x)(1-y)}\right]\Phi_n(y,r) \, , \label{com1}
\end{eqnarray}
as discussed in Refs.\ \cite{Callan:ps,Burkardt:2000ez}.
Multiplying this by $\Phi_n(x,r)$ and integtating over $x$ gives
zero on the l.h.s., and hence the relation  
\begin{eqnarray}
m_{01}^2\left(\int_0^1dz
\frac{\Phi_n(z,r)}{z}\right)^2 = m_{02}^2\left(\int_0^1dz
\frac{\Phi_n(z,r)}{1-z}\right)^2\, .
\end{eqnarray}
Taking the square root of both sides of this equation gives
\begin{eqnarray}
m_{01}\int_0^1dz
\frac{\Phi_n(z,r)}{z} =\pm m_{02} \int_0^1dz
\frac{\Phi_n(z,r)}{1-z}\equiv C_n\, , \label{idet4}
\end{eqnarray}
where the sign is positive if the phase of $\Phi_n(0,r)$ is the
same as $\Phi_n(1,r)$ and negative if it is opposite. 
Since the states are non-degenerate, the phase of the $n$th
eigenstate is $(-1)^{(n-1)}$. 
Combining this with Eq.\ (\ref{B6}) gives    
\begin{eqnarray}
M_n^2\int_0^1dz\,\Phi_n(z,r)= \left[m_{01}+(-1)^{(n-1)} m_{02}
\right] C_n\, , \label{B10}
\end{eqnarray}
where the constant $C_n$ is yet to be determined.  In this paper
we have chosen the sign of $\Phi_n(z,r)$ so that it is always
positive as $z\to0$, so that $C_n>0$ for all $n$.

\begin{figure}
\leftline{
\mbox{\hspace{-0.4in}
   \epsfysize=2.5in\epsffile{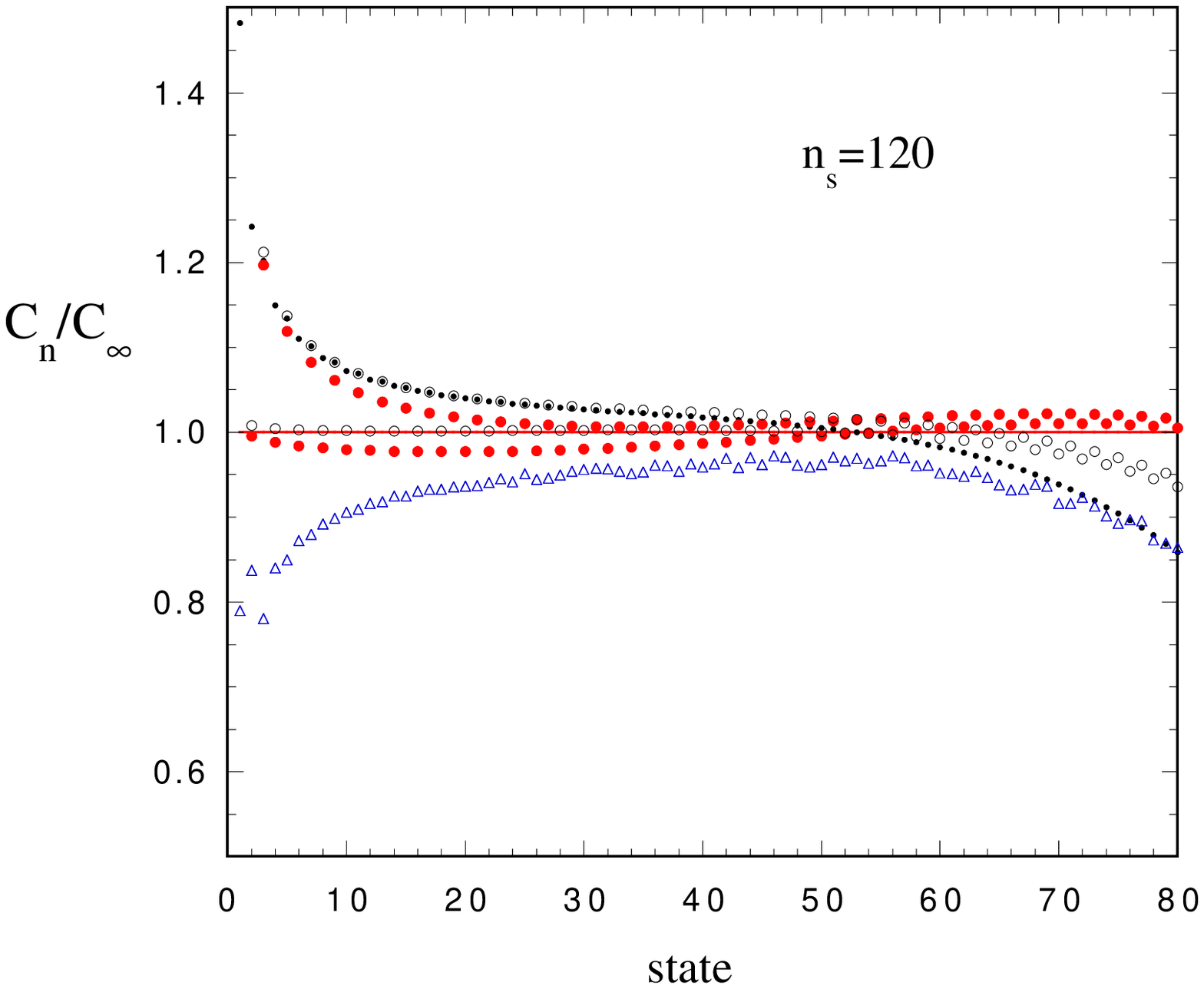}
}
}
\vspace*{-2.5in}
\hspace*{0.4in}
\rightline{
\mbox{
  \epsfysize=2.5in\epsffile{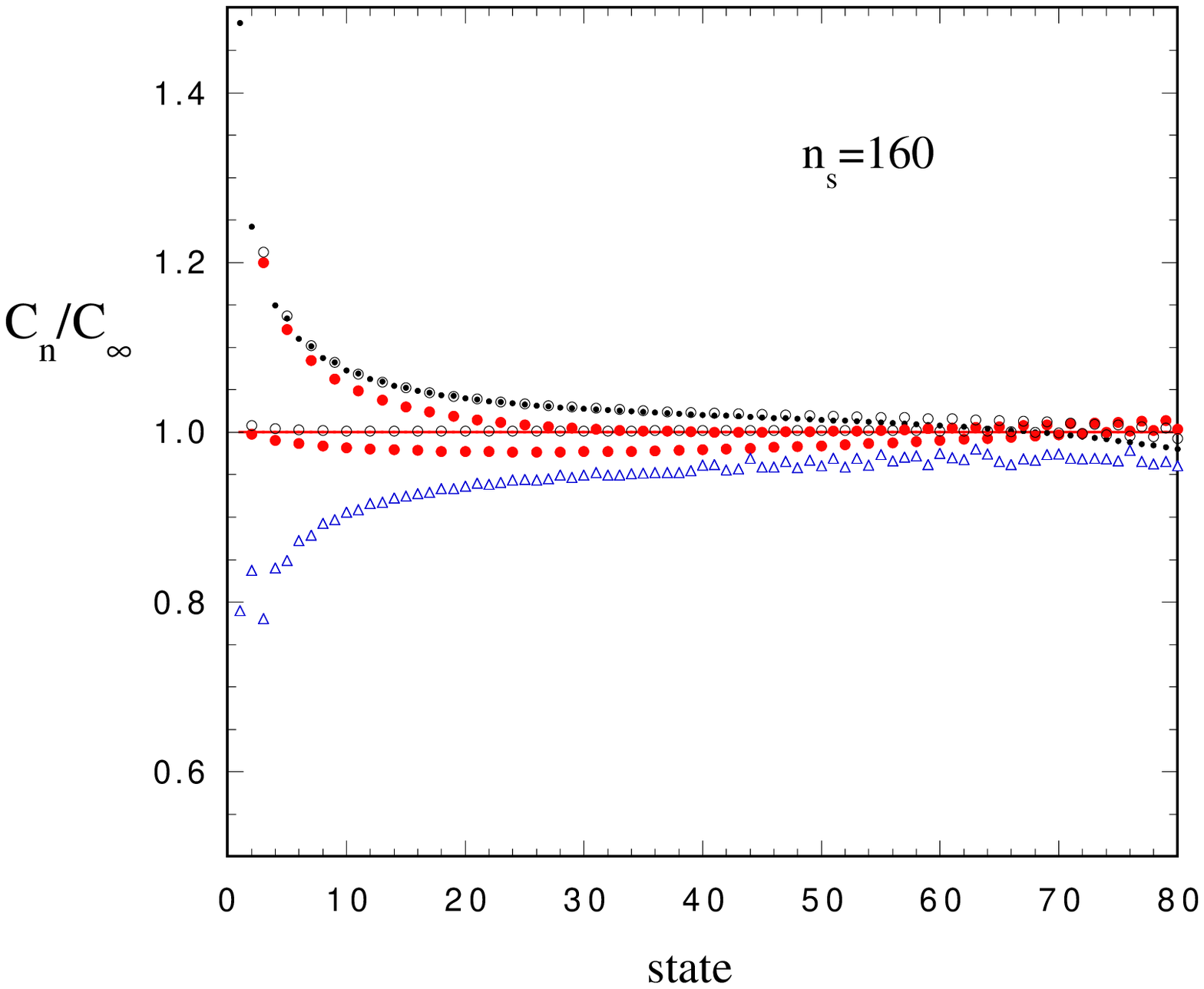}
}
}
\caption{\footnotesize\baselineskip=10pt Numerical test of the relations (\ref{idet4}),
(\ref{identB}), and (\ref{B11}) for $m_{01}=1.5$ and $m_{02}=3$
and different numbers of splines, $n_s$.  Solid circles are values
of the l.h.s. of (\ref{idet4}) divided by
$C_\infty$; open circles are the absolute value of the r.h.s.
divided by $C_\infty$.  They should be equal for each state, and
approach unity as $n\to\infty$.  The tiny solid squares are the
values of $(\mu_{n+1}^2-\mu_n^2)/\pi^2$ which, according
(\ref{B11}),  approach unity as $n\to\infty$.  The open triangles
are the square of the normalization constant divided by 2 (for
wave functions initially normalized with their maximum value
equal to unity), and approaches unity if the approximation
(\ref{ansatz}) is valid. }
\label{fig:idet1}
\end{figure}

Our demonstration of duality depends upon the relations
\begin{eqnarray}
&&\lim_{n\to\infty} C_n \to C_\infty = g
\sqrt{
\pi}
\label{identB}\\ 
&&\lim_{n\to\infty}
\left(M_{n+1}^2-M_n^2\right)\to 
C_\infty^2=\pi g^2 \, ,
\label{B11}
\end{eqnarray}
which can be derived approximately using arguments
given by 't Hooft \cite{thooft}, and Callan, Coote, and Gross
\cite{Callan:ps}, given here for completeness.  First,
for large $n$, the boundary conditions require that the normalized
wave function go like    
\begin{eqnarray}
\Phi_n(z,r) \sim  
\sqrt{2}\,\sin(n\pi z) \, .
\label{ansatz}
\end{eqnarray}
One test of this approximation is shown in Fig. \ref{fig:idet1}.
If $n$ is very large, so that $M_n\gg m_{0i}$, the behavior of the
wave function at the endpoints can be ignored, and near $z=1/2$
Eq.~(\ref{2eqee}) can be approximated 
\begin{eqnarray}
M_n^2\,\Phi_n(z,r)\simeq
-\frac{g^2}{\pi}\dashint_{-\infty}^\infty  
\frac{\Phi'_n(y,r)}{y-z}\, .
\end{eqnarray}
Substituting the ansatz (\ref{ansatz}) into this equation gives 
\begin{eqnarray}
M_n^2\,\sin(n\pi z)&=&
-g^2 n \dashint_{-\infty}^\infty  
\frac{\cos(n\pi y)}{y-z} = - g^2 n \, {\rm Re}
\int_{-\infty}^\infty \frac{\cos(n\pi y)}{y-z-i\epsilon}
\nonumber\\
&=& \pi g^2 n \,\sin(\pi n z) \, .
\end{eqnarray}
Hence $M_n^2\to \pi g^2 n$ as $n\to\infty$ and identity
(\ref{B11}) is proved.  To prove (\ref{identB}) we consider
the matrix element of (\ref{com1}) 
\begin{eqnarray}
\int_0^1\Phi_{n+1}(x,r)\,[H,{\cal K}]\,\Phi_n(x,r)&=& 2C_{n+1}C_n 
\nonumber\\
&=&\left(M_{n+1}^2-M_n^2\right)\dashint_0^1 dxdy
\frac{\Phi_{n+1}(x,r)\Phi_n(y,r)}{y-x}\, .
\end{eqnarray}
where we used the fact that the phases of $\Phi_n$ and
$\Phi_{n+1}$ are opposite.  The integral is evaluated in the
large $n$ approximation 
\begin{eqnarray}
\dashint_0^1 dxdy
\frac{\Phi_{n+1}(x,r)\Phi_n(y,r)}{y-x} &\simeq& 2 
\;{\rm Re}
\int_0^1dx\int_{-\infty}^\infty  dy\frac{\sin([n+1]\pi
x)\sin(n\pi y)}{y-x-i\epsilon} \nonumber\\
&=&2 
\int_0^1 dx \sin([n+1]\pi x)\cos(n\pi
x)\nonumber\\ &=& 2 
\left(1 +\frac{1}{2n+1}\right)  
\to 2 
\, .
\end{eqnarray}
Hence, 
\begin{eqnarray}
C_{n+1}C_n\simeq 
\pi g^2\simeq C_\infty^2 
\end{eqnarray}
and (\ref{identB}) is proved.

A numerical demonstration of the identities (\ref{idet4}), 
(\ref{identB}), and (\ref{B11}) is given in Figs.\
\ref{fig:idet1}.  The identities are satisfied to better than a
few percent for state numbers between 40 and 80, provided the
number of splines is at least 160 (twice the number of the
maximum state of interest).    

\begin{figure}
\leftline{
\mbox{\hspace{-0.4in}
   \epsfysize=2.5in\epsffile{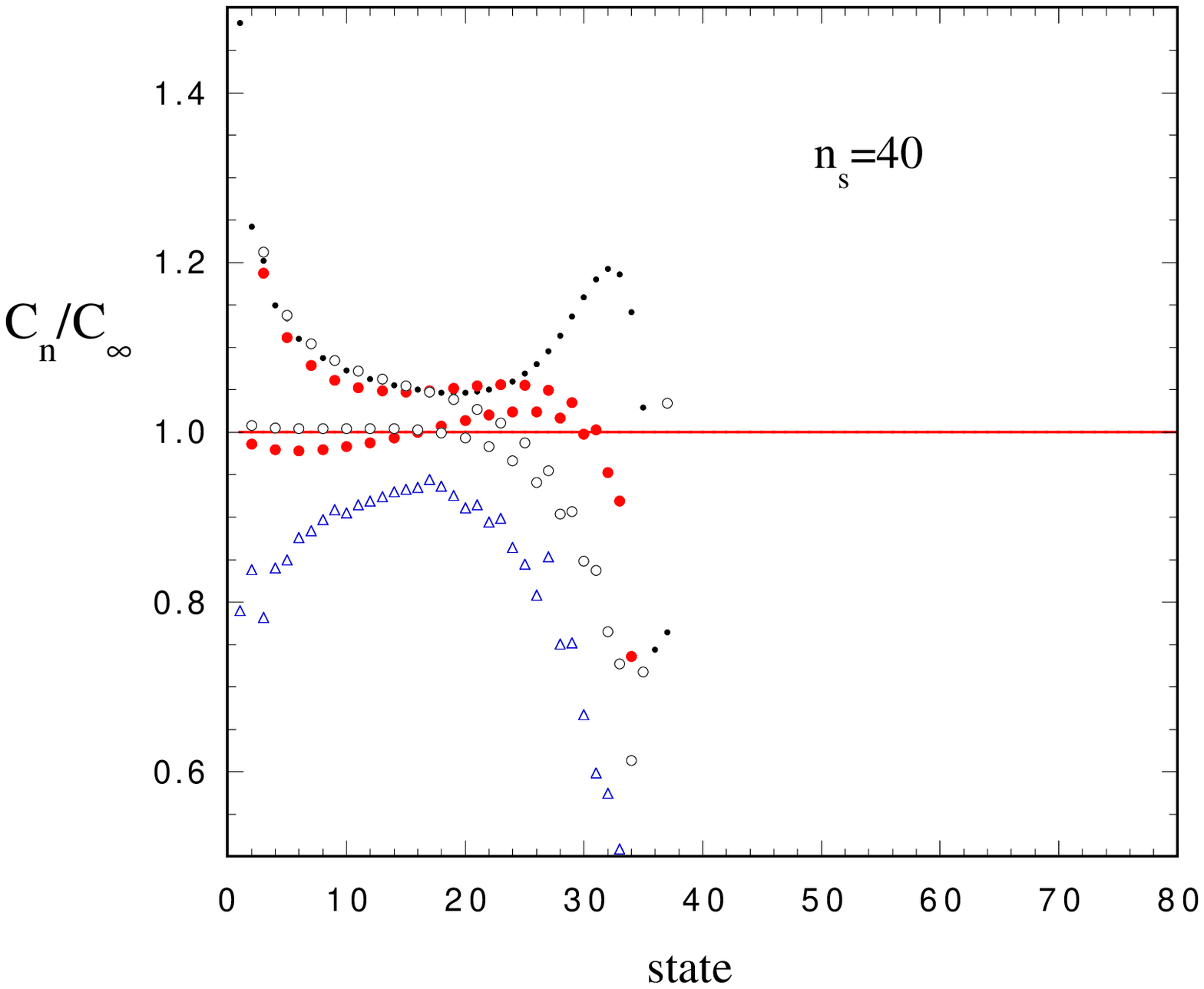}
}
}
\vspace*{-2.5in}
\hspace*{0.4in}
\rightline{
\mbox{
   \epsfysize=2.5in\epsffile{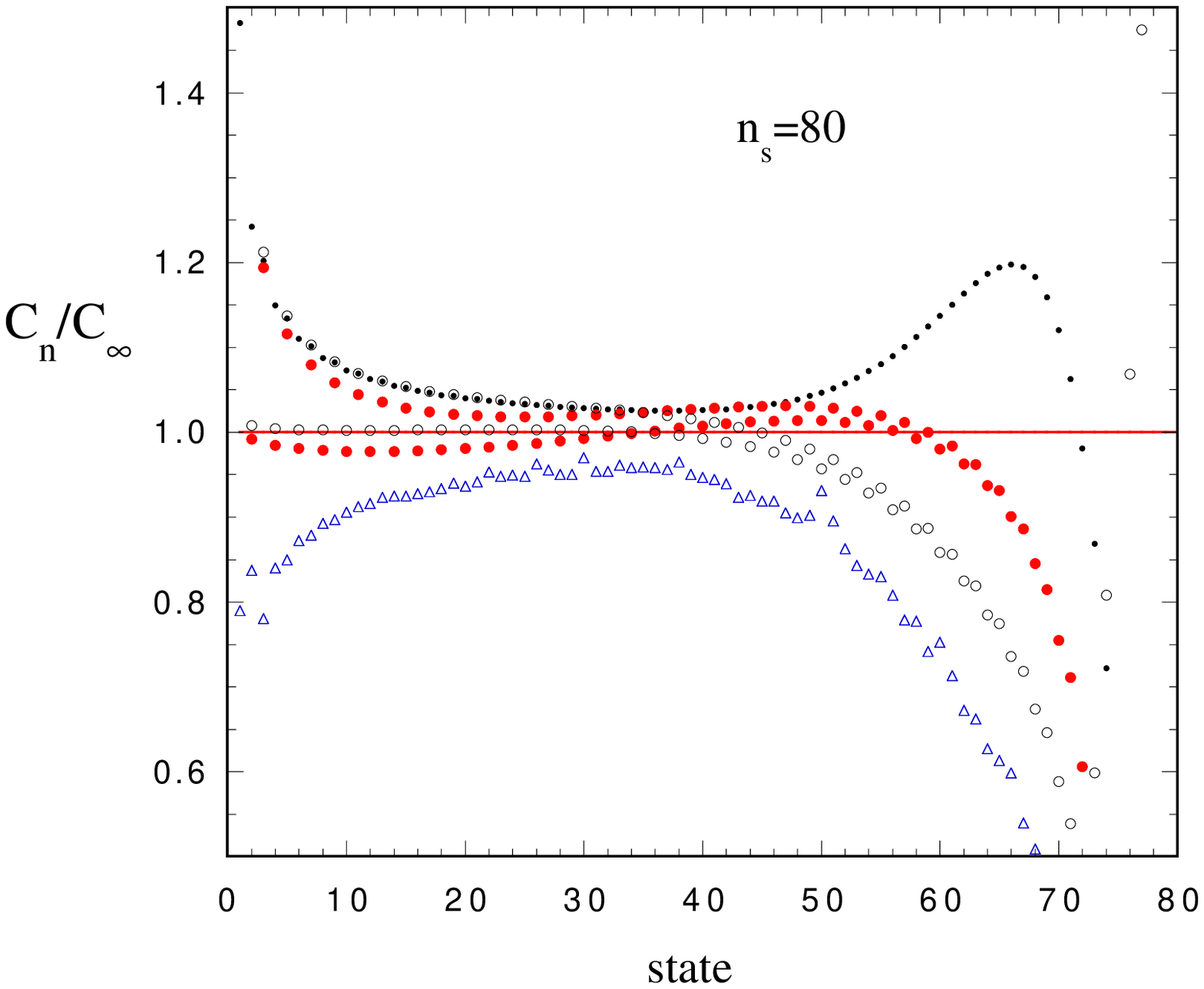}
}
}
\end{figure}

Two additional relations that follow from
the completeness relation (\ref{complete}) and the bound state
equation are needed.  If $z\in [0,1]$, then:
\begin{eqnarray}
&&
\sum_n\,\Phi_n(z,r_n)\int_0^1dz'\,
\Phi_n(z',r_n)= 1 \label{id1}\\ 
&&\sum_n\left(\frac{m_{01}^2}{z}+\frac{m_{02}^2}{1-z}-M_n^2
\right)\,\Phi_n(z,r_n)\int_0^1dz'\,
\Phi_n(z',r_n)=0\label{id2}
\, .
\end{eqnarray}
The proof of the second relation (\ref{id2}) follows from
application of the bound state equation
\begin{eqnarray}
&&\sum_n\left(\frac{m_{01}^2}{z}+\frac{m_{02}^2}{1-z}-M_n^2
\right)\,\Phi_n(z,r_n)\int_0^1dz'\,
\Phi_n(z',r_n)
\nonumber\\
&&\qquad=\sum_n\int_0^1dy\,\left[\frac{\Phi_n(y,r_n)-
\Phi_n(z,r_n)} {(y-z)^2} \right]\int_0^1dz'\,
\Phi_n(z',r_n)\nonumber\\
&&\qquad=\sum_n\left\{\int_0^{z-\epsilon} +
\int_{z-\epsilon}^{z+\epsilon}+\int_{z+\epsilon}^1\right\} dy\,
\left[\frac{\Phi_n(y,r_n)-\Phi_n(z,r_n)} {(y-z)^2}
\right]\int_0^1dz'\,
\Phi_n(z',r_n)\nonumber\\
&&\qquad\to\sum_n\frac{1}{2}\int_{z-\epsilon}^{z+\epsilon}dy\;
\Phi_n''(z,r_n)\int_0^1dz'\,\Phi_n(z',r_n) =
\nonumber\\ &&\qquad= \epsilon\,\sum_n
\Phi_n''(z,r_n)\int_0^1dz'\,\Phi_n(z',r_n)\to0
\, ,
\end{eqnarray}
where the identity (\ref{id1}) is used to eliminate the
nonsingular integrals over the intervals $[0,z-\epsilon]$ and
$[z+\epsilon,1]$, and the singular part of the integral is
proportional to $\epsilon$ which vanishes as $\epsilon\to0$.

If Eq.~(\ref{phiout}) is used, modifications of the completeness
relation (\ref{complete}) and the identity (\ref{id1}), which
hold in the region $z\notin[0,1]$, can be derived.  The results
are 
\begin{eqnarray}
&&
\sum_n\left(\frac{m_1^2}{z}+\frac{m^2_2}{1-z}-
M_n^2\right)
\Phi_n(z,r_n)\Phi_n(z',r_n) =
\frac{g^2/\pi}{(z-z')^2}\qquad z'\in[0,1]
\label{complete2}\\
&&
\sum_n\left(\frac{m_1^2}{z}+\frac{m^2_2}{1-z}-
M_n^2\right)\Phi_n(z,r_n)\int_0^1dz'\,
\Phi_n(z',r_n)= -\frac{g^2/\pi}{z(1-z)} \label{id3} 
\, .
\end{eqnarray}

Note that the completeness relation and all of these identities
hold for {\it any\/} two quark Hamiltonian.  In particular, the
Hamiltonian may describe bound states with flavor (where, in
general,
$m_1\ne m_2$) as well as flavorless bound states where
$m_1=m_2$.  In some applications we will study the
coupling of a photon to a $q\bar q$ pair.  Such a coupling
necessarily involves {\it flavorless\/} states only, and we
will use the identities for the flavorless states.

\section{Summation of the quark scattering series}
\label{appen:B}

Here we show how to sum this series (\ref{eq249})
and obtain a useful form for $M$.
 
Since $V$ is independent of
the plus components of momentum, the integrations over the plus
components of momentum can be carried out.  Using the identity
(\ref{2eqab}) [with $q_-<0$ in this case], the first term on
the rhs of (\ref{eq249}) becomes
\begin{eqnarray}
\left<MG{\cal O}\right>_1(p';q)
&=&i\int\,\frac{d^2p}
{\pi^2}
\frac{V(p',p)\,{\cal O}(p_-,q) }
{d_2(p_-,p_+,q)}\nonumber\\
&=&  
\int_0^{1}{dz}\,{{\rm V}(z',z)}\;
\frac{{\cal O}(p_-,q)}{\Delta(z,q^2)}
-\frac{g^2}{\pi\,z'(1-z')}\frac{{\cal
O}(p'_-,q)}{\Delta(z',q^2)}\nonumber\\
&=&\int_0^{1}{dz} \,\left\{{\rm V}(z',z)-
\frac{g^2\,\delta(z'-z)}{\pi\,z(1-z)}
\right\}\frac{{\cal O}(p_-,q)}{\Delta(z,q^2)}
\quad{\rm if}\; z'\in[0,1]\, ,\quad\quad\quad\label{first}
\end{eqnarray}
or
\begin{eqnarray}
\left<MG{\cal O}\right>_1(p';q)
&=&i\int\,\frac{d^2p}
{\pi^2}
\frac{V(p',p)\,{\cal O}(p_-,q) }
{d_2(p_-,p_+,q)}
=\int_0^{1}{dz} \,{\rm V}_0(z',z)\frac{{\cal
O}(p_-,q)}{\Delta(z,q^2)}
\quad{\rm if}\; z'\notin[0,1]\,
.\quad\quad\quad\label{firsta}
\end{eqnarray}
Here the replacement (\ref{eq251a}) has been used to convert
$V\to{\rm V}$ (or V$_0$). Similarly, the second term becomes 
\begin{eqnarray}
\left<MG{\cal O}\right>_2(p';q)
&=&i\int\,\frac{d^2k}
{\pi^2}
\frac{V(p',k)}{d_2(k_-,k_+,q)}
\int_0^{1}{dz}\,\left\{{\rm V}(y,z)-
\frac{g^2\,\delta(y-z)}{\pi\,z(1-z)}
\right\}\frac{{\cal O}(p_-,q)}{\Delta(z,q^2)}\nonumber\\
&=&\int_0^{1}{dy}\,\left\{{\rm V}(z',y)-
\frac{g^2\,\delta(z'-y)}{\pi\,y(1-y)}
\right\}\,\frac{1}{\Delta(y,q^2)} \nonumber\\
&&\qquad\times\int_0^{1}{dz}\,\left\{{\rm
V}(y,z)-
\frac{g^2\,\delta(y-z)}{\pi\,z(1-z)}
\right\}\frac{{\cal O}(p_-,q)}{\Delta(z,q^2)}
\quad{\rm if}\; z'\in[0,1]
\, ,\quad\quad\quad \label{second}
\end{eqnarray}
or
\begin{eqnarray}
\left<MG{\cal O}\right>_2(p';q)
&=&\int_0^{1}{dy}\,{\rm V}_0(z',y)\,\frac{1}{\Delta(y,q^2)}
\nonumber\\ &&\qquad\times\int_0^{1}{dz}\,\left\{{\rm
V}(y,z)-
\frac{g^2\,\delta(y-z)}{\pi\,z(1-z)}
\right\}\frac{{\cal O}(p_-,q)}{\Delta(z,q^2)}
\quad{\rm if}\; z'\notin[0,1]
\, ,\quad\quad\quad \label{seconda}
\end{eqnarray}
and the third term follows the same pattern.  Replacing
$(q_-)^2M(p',p;q)/\pi\to{\rm M}(z',z;q^2)$ and ${\cal
O}(p_-,q)\to{\rm O}(z,q^2)$, the series (\ref{eq249}), in the
interval $z'\in[0,1]$, is clearly summed by the following
equation
\begin{eqnarray}
&&\int_0^1{dz}\,{{\rm
M}(z',z;q^2)}\,\frac{{\rm O}(z,q^2)}{\Delta(z,q^2)}
=\int_0^{1}{dz} \,\left\{{\rm V}(z',z)-
\frac{g^2\,\delta(z'-z)}{\pi\,z(1-z)}
\right\}\frac{{\rm O}(z,q^2)}{\Delta(z,q^2)}
\nonumber\\
&&\qquad+\int_0^{1}{dy}
\,\frac{{\rm M}(z',y ;q^2)}{\Delta(y,q^2)}
\,\int_0^{1}{dz}\,
\left\{{\rm V}(y,z)-
\frac{g^2\,\delta(y-z)}{\pi\,z(1-z)}
\right\}\frac{{\rm O}(z,q^2)}{\Delta(z,q^2)}
\quad{\rm if}\; z'\in[0,1]
\, .\qquad\qquad \label{eq253}
\end{eqnarray}
Introducing the shorthand notation 
\begin{eqnarray}
F(z',z;q^2)=\frac{{\rm M}(z',z;q^2)}{\Delta(z,q^2)}\, ,
\end{eqnarray}
Eq.~(\ref{eq253}) can be rearranged into the following form 
\begin{eqnarray}
&&\int_0^1\,dz\,F(z',z;q^2)\,\left\{\left(\Delta(z,q^2)+
\frac{g^2/\pi}{z(1-z)}\right)\,\frac{{\rm
O}(z,q^2)}{\Delta(z,q^2)} -\int_0^1dy\,{\rm V}(z,y)\,\frac{{\rm
O}(y,q^2)}{\Delta(y,q^2)}
\right\}\qquad
\qquad\nonumber\\
&&\qquad\qquad=\int_0^1\,dz\,F(z',z;q^2)\,
\left(H(z)-q^2\right)\,\frac{{\rm
O}(z,q^2)}{\Delta(z,q^2)}\nonumber\\
&&\qquad\qquad=\int_0^{1}{dz} \,\left\{{\rm V}(z',z)-
\frac{g^2\,\delta(z'-z)}{\pi\,z(1-z)}
\right\}\frac{{\rm O}(z,q^2)}{\Delta(z,q^2)} \, ,\qquad
\label{eq255}
\end{eqnarray}
which requires that $F$ satisfy the equation  
\begin{eqnarray}
F(z',z,q^2)\,\left(H(z)-q^2\right)
&=&{\rm V}(z',z)-
\frac{g^2\,\delta(z'-z)}{\pi\,z(1-z)}
\, .\quad \label{eq257}
\end{eqnarray}
This equation is easily solved with the use of the Greens
function:
\begin{eqnarray}
F(z',z;q^2)&=&-\int_0^1dy\,\left\{{\rm V}(z',y)-
\frac{g^2\,\delta(z'-y)}{\pi\,y(1-y)}
\right\}\,G(y,z,q^2)
\nonumber\\
&=&-\int_0^1dy\,\left\{{\rm V}(z',y) -
\frac{g^2\,\delta(z'-y)}{\pi\,y(1-y)} \right\}
\sum_n\,\frac{
\Phi_n(y,r_n)\Phi_n(z,r_n)}{q^2-M_n^2}
\nonumber\\
&=&
\sum_n\left(M_n^2-\frac{m_{1}^2}{z'}-
\frac{m_{2}^2}{1-z'}\right)
\frac{\Phi_n(z',r_n)\Phi_n(z,r_n)}
{q^2-M_n^2}
\, .\quad
\label{eq258}
\end{eqnarray}
Putting it all together gives Eq.\ (\ref{eq267}). 
 
If $z'\notin[0,1]$, the series reduces to the same formal
result
\begin{eqnarray}
\int_0^1dz\,{\rm M}(z',z;q^2)\,\frac{{\rm
O}(z,q^2)}{\Delta(z,q^2)}&=&\int_0^{1}{dz}  \,{\rm
V}_0(z',z)\frac{{\rm O}(z,q^2)}{\Delta(z,q^2)}
\nonumber\\
&+&\int_0^{1}{dy} \,{\rm V}_0(z',y)\frac{1}{\Delta(y,q^2)}
\int_0^1dz\,{\rm M}(y,z;q^2)\,\frac{{\rm
O}(z,q^2)}{\Delta(z,q^2)}
\nonumber\\
&=&\int_0^{1}{dz}  \,{\rm
V}_0(z',z)\frac{{\rm O}(z,q^2)}{\Delta(z,q^2)} +\sum_n\int_0^{1}{dy} \,{\rm
V}_0(z',y)
\nonumber\\
&&\qquad\times\frac{\Delta(y,M_n^2)}{\Delta(y,q^2)}
\frac{
\Phi_n(y,r_n)}{M_n^2-q^2}
\int_0^1dz\,\Phi_n(z,r_n) \,{\rm O}(z,q^2)\nonumber\\
&=&
\sum_n\Delta(z',M_n^2)\,
\frac{\Phi_n(z',r_n)}{M_n^2-q^2}
\int_0^1dz\,\Phi_n(z,r_n) \,{\rm O}(z,q^2)
\, ,\qquad\qquad \label{eq267a}
\end{eqnarray}
where the completeness relation (\ref{complete}) and
Eq.~(\ref{phiout}) were used in the last step.  This shows
that (\ref{eq267}) can be applied for {\it any\/}
$z'\in[-\infty,\infty]$.

\section{summation of the quark current series}
\label{appen:C}  

The series (\ref{current1a}) for the dressed quark current is
evaluated in this Appendix.  The ``first'' integral over
$k_+$ in each term in the series is evaluated as in
Eq.~(\ref{first}) or (\ref{firsta}) 
\begin{eqnarray}
&&i\int\frac{d^2k}{(2\pi)^2}\,V(p',k)\,
\gamma_-\,S_i(k)\gamma^\mu S_i(k-q)\,\gamma_-\equiv
i\int\frac{d^2k}{\pi^2}\,\frac{V(p',k)\,
N_i^\mu(k_-,q_-)}{d^i_2(k_-,k_+,q)}\nonumber\\
&&\qquad\qquad=\cases{{\displaystyle\int_0^1dy\,\left\{{\rm
V}(z',y)-
\frac{g^2\,\delta(z'-y)}{\pi\,y(1-y)}
\right\}\frac{N_i^\mu(k_-,q_-)}{\Delta_i(y,q^2)} }& if
$z'\in[0,1]$\cr 
{\displaystyle\int_0^1dy\,{\rm
V}_0(z',y)\,\frac{N_i^\mu(k_-,q_-)}{\Delta_i(y,q^2)} }& if
$z'\notin[0,1]\, ,\qquad\quad$}
\label{current1b}
\end{eqnarray}
where  
\begin{eqnarray}
N_i^\mu (k_-,q_-)=\frac{
\gamma_-[m_{0i}+k_-\gamma_+]\,\gamma^\mu\,\left[m_{0i} +
(k_--q_-)\gamma_+\right]\,\gamma_-}{4\,k_-(k_--q_-)}\, ,
\label{current1bb}
\end{eqnarray}
and $d_2^i$ and $\Delta_i$ are (\ref{d2}) and (\ref{d1})
with equal masses $m_1=m_2=m_i$.  Note that the only dependence
of the integrand on $k_+$ was in the denominator $d_2^i$, so the
$k_+$ integral could be evaluated using the methods of Appendix
\ref{appen:B}.  The second integral is similarily evaluated.  If
$z'\in[0,1]$, the following result is obtained for the series
Eq.~(\ref{current1a}) 
\begin{eqnarray}
&&j_i^\mu(p',p'-q)=j_i^\mu(z',q^2)=\gamma^\mu
+\int_0^{1}{dy} \,\left\{{\rm V}(z',y)-
\frac{g^2\,\delta(z'-y)}{\pi\,y(1-y)}
\right\}\frac{{\rm N}^\mu_i(y,q)}{\Delta_i(y',q^2)}
\nonumber\\
&&\quad+\int_0^{1}{dy'}\left\{{\rm V}(z',y')-
\frac{g^2\,\delta(z'-y')}{\pi\,y'(1-y')}
\right\}
\,\frac{1}{\Delta_i(y',q^2)}
\,\int_0^{1}{dy}\,
\left\{{\rm V}(y',y)-
\frac{g^2\,\delta(y'-y)}{\pi\,y(1-y)}
\right\}\frac{{\rm N}^\mu_i(y,q)}{\Delta_i(y,q^2)}
\nonumber\\
&&\quad+\cdots\quad=\gamma^\mu+\int_0^1{dz}\,{{\rm
M}(z',y;q^2)}\,\frac{{\rm N}^\mu_i(y,q)}{\Delta_i(y,q^2)}\, ,
\label{current2}
\end{eqnarray}
where ${\rm N}^\mu_i(y,q)={N}^\mu_i(k_-,q_-)$, and the series
is summed using Eq.~(\ref{eq253}) for the scattering matrix
M.  Using the solution (\ref{eq267}) for the scattering matrix
the result, for all values of $z'=p'_-/q_-$, is  
\begin{eqnarray}
j_i^\mu(z',q^2)=\gamma^\mu +
\sum_n\Delta_i(z',M_n^2)\,
\frac{\Phi_n(z',r_n)}{M_n^2-q^2}
\,\int_0^1dz\,\Phi_n(z,r_n) \,{\rm N}^\mu_i(z,q)
\, ,\qquad\qquad \label{current2a}
\end{eqnarray}
where $j_i^\mu(p',p'-q)=j_i^\mu(z',q^2)$, and $\Delta_i$ is the
$\Delta$ of Eq.\ (\ref{d1}) with equal masses $m_1=m_2=m_i$. 
 
The same expression also holds for $z'\notin[0,1]$, but the
derivation differs.  As before, the full result can be deduced 
from the form of the first two terms of the series
(\ref{current1a}).  Guided by (\ref{current1b}),  
and using the steps already displayed
in Eq.~(\ref{eq267a}), we obtain  
\begin{eqnarray}
&&j_i^\mu(z',q^2)=\gamma^\mu
+\int_0^{1}{dy} \,{\rm V}_0(z',y)
\,\frac{{\rm N}^\mu_i(y,q)}{\Delta_i(y',q^2)}
\nonumber\\
&&\quad+\int_0^{1}{dy'}\,{\rm V}_0(z',y')
\,\frac{1}{\Delta_i(y',q^2)}
\,\int_0^{1}{dy}\,
\left\{{\rm V}(y',y)-
\frac{g^2\,\delta(y'-y)}{\pi\,y(1-y)}
\right\}\frac{{\rm N}^\mu_i(y,q)}{\Delta_i(y,q^2)}
\nonumber\\
&&\quad+\cdots\quad=\gamma^\mu+\int_0^1{dz}\,{{\rm
M}(z',y;q^2)}\,\frac{{\rm N}^\mu_i(y,q)}{\Delta_i(y,q^2)}
\nonumber\\
&&\quad=\gamma^\mu   +
\sum_n\Delta_i(z',M_n^2)\,
\frac{\Phi_n(z',r_n)}{M_n^2-q^2}
\,\int_0^1dz\,\Phi_n(z,r_n) \,{\rm N}^\mu_i(z,q)\, .
\label{current2ab}
\end{eqnarray}
We see, as we did with
Eq.~(\ref{eq267}), that (\ref{current2a}) holds for {\it
all\/} $z'$.

Equation (\ref{current2a}) expresses the dressed current as a
sum over contributions from the bound states that couple to the
photon, showing that vector dominance is a rigorous
consequence of this model.

The structure of the numerator N$^\mu_i$ depends on the bare
current: 
\begin{eqnarray}
&\gamma_-\,:\qquad&{\rm N}^-_i(y,q) =\,\gamma_-\nonumber\\
&\gamma_+\,:\qquad&{\rm N}^+_i(y,q)
=\,-\gamma_-\,\frac{m^2_{0i}}{2\,q_-^2\,y(1-y)}
\nonumber\\
&\gamma_x\,:\qquad&{\rm N}^x_i(y,q)
=\,-\gamma_x\gamma_-\,\frac{m_{0i}}{ 2\,q_-\,y(1-y)}
\, .
\label{current3}
\end{eqnarray}
Note the interesting structure of the correction term to the
bare current
$\gamma_x$, which will be of central importance in our
discussion below.  Substituting these results into
(\ref{current2a}) gives the following

\begin{eqnarray} 
j_{i-}(z',q^2)&=&\gamma_-\left\{1 + 
\sum_n\Delta_{i}(z',M_n^2)
\frac{\Phi_n(z',r_n)}{M_n^2-q^2}\;
\int_0^1dy\;\Phi_n(y,r_n)\,
\right\}\nonumber\\
j_{i+}(z',q^2)&=&\gamma_+ -
\gamma_-\,\frac{m_{0i}^2}{2q_-^2} \,\left\{
\sum_n\Delta_{i}(z',M_n^2)
\frac{\Phi_n(z',r_n)}{M_n^2-q^2}\;
\int_0^1dy\,\frac{\Phi_n(y,r_n)}{y(1-y)}\right\} \nonumber\\
j_x(z',q^2)&=&\gamma_x  -
\gamma_x\,\gamma_-\,\,\frac{m_{0i}}{2q_-} \,\left\{
\sum_n\Delta_{i}(z',M_n^2)
\frac{\Phi_n(z',r_n)}{M_n^2-q^2}\;
\int_0^1dy\,\frac{\Phi_n(y,r_n)}{y(1-y)}\right\}\, .
\label{current5}
\end{eqnarray}

We organize these expressions by introducing the quark form
factor, defined in Eq.\ (\ref{ff1}).  This immediately gives Eq.\
(\ref{ff3aa}) for the  $j_-$ current. The 
other currents in (\ref{current5}) can also be written in terms of
$F_i$.  To this end note that the bound state equation for
equal mass quarks implies that 
\begin{eqnarray} 
M_n^2 \int_0^1 dy\, \Phi_n(y,r_n)=m_{0i}^2\int_0^1 dy\;
\frac{\Phi_n(y,r_n)}{y(1-y)}\, .
\label{cond1}
\end{eqnarray}
Hence the quark mass can be removed from the second term in
$j_+$, giving   
\begin{eqnarray} 
g_i(z',q)&\equiv&-\frac{m_{0i}^2}{2q_-^2} \,\left\{
\sum_n\Delta_{i}(z',M_n^2)
\frac{\Phi_n(z',r_n)}{M_n^2-q^2}\;
\int_0^1dy\,\frac{\Phi_n(y,r_n)}{y(1-y)}\right\}
\nonumber\\
&=&\frac{q_+}{q_-}\,\left\{
\sum_n\frac{M_n^2 }{Q^2}\,\Delta_{i}(z',M_n^2)\,
\frac{\Phi_n(z',r_n)}{M_n^2+Q^2}\;
\int_0^1dy\,\Phi_n(y,r_n)\right\}\nonumber\\
&=&-\frac{q_+}{q_-}\,F_i(z',Q^2)+\frac{q_+
}{q_-\,Q^2}\,
\left\{\sum_n\Delta_{i}(z',M_n^2)\,
\Phi_n(z',r_n)\;
\int_0^1dy\,\Phi_n(y,r_n)\right\}\, ,\qquad
\label{ff2}
\end{eqnarray}
where $g_i(z',q)$ defined by this equation should not be
confused with the two body Greens function or the bound state
vertex function.  If
$z'\in[0,1]$, the second term can be reduced using the
identities (\ref{id1}) and (\ref{id2}) 
\begin{eqnarray} 
&&
\sum_n\left[\frac{m_{0i}^2}{z'(1-z')}-M_n^2-
\frac{g^2/\pi}{z'(1-z')}\right]
\Phi_n(z',r_n)\;
\int_0^1dy\,\Phi_n(y,r_n)
=-\frac{g^2}{\pi}\frac{1}{z'(1-z')}
\, .\qquad\quad
\label{ff3}
\end{eqnarray}
The same result is obtained directly from identity (\ref{id3})
if $z'\notin[0,1]$. Hence, for all $z'$,
\begin{eqnarray} 
g_i(z',q)&=&-\frac{q_+}{q_-}\,F_i(z',Q^2)  +
\frac{g^2}{2\pi}\frac{1}{q_-^2\,z'(1-z')}=
-\frac{q_+}{q_-}\,F_i(z',Q^2)  -
\frac{g^2}{2\pi}\frac{1}{p'_-p_-}
\, .\qquad
\label{ff3a}
\end{eqnarray}

Finally, the $j_x$ component of the current can be similarly
reduced.  The second term in $j_x$ is proportional to
\begin{eqnarray}  
H_i(z',q)&\equiv&-\frac{m_{0i}}{2q_-}\,\left\{
\sum_n\Delta_{i}(z',M_n^2)
\frac{\Phi_n(z',r_n)}{M_n^2-q^2}\;
\int_0^1dy\,\frac{\Phi_n(y,r_n)}{y(1-y)}\right\}
\nonumber\\
&=&-\frac{1}{2q_-\,m_{0i}}\left\{
\sum_nM_n^2\,
\Delta_{i}(z',M_n^2)
\frac{\Phi_n(z',r_n)}{M_n^2+Q^2}\;
\int_0^1dy\,\Phi_n(y,r_n)\right\}\nonumber\\
&=&-\frac{q_+}{m_{0i}}\,F_i(z',Q^2) - \frac{
}{2q_-\,m_{0i}}
\left\{\sum_n
\Delta_{i}(z',M_n^2)\,
\Phi_n(z',r_n)\;
\int_0^1dy\,\Phi_n(y,r_n)\right\}\, .\qquad\quad
\end{eqnarray}
The part of the second term in $\{\}$ is the same constant
derived above, and combining the results gives Eq.\ (\ref{ff6}).


\section{Kinematics for DIS}
\label{appen:E}

Kinematics for DIS in the c.m. frame are given in this Appendix.
When $Q^2$ is very large, the components of $P$
and $q$ in this frame are (see Ref.~\cite{ref1})
\begin{eqnarray}
q_+=&&Q\sqrt{\frac{1-x}{2x}}
\left[1-\frac{M_0^2 x (1-2x)}{2Q^2(1-x)}+...\right]    
\nonumber\\
q_-=&&-Q\sqrt{\frac{x}{2(1-x)}}
\left[1+\frac{M_0^2 x (1-2x)}{2Q^2(1-x)}+...\right] 
\nonumber\\ 
P_+=&&\frac{M_0^2}{Q}\sqrt{\frac{x(1-x)}{2}}
\left[1-\frac{M_0^2x(1-2x^2)}{2Q^2(1-x)}+...\right]
\nonumber\\ 
P_-=&&\frac{Q}{\sqrt{2x(1-x)}}
\left[1+\frac{M_0^2x(1-2x^2)}{2Q^2(1-x)}+...\right]\, .
\label{5eq5}
\end{eqnarray}
The energy and momentum of particles 1 and 2 in the final state
are
\begin{eqnarray}
p_z&=&\pm \frac{Q}{2}\sqrt{\frac{1-x}{x}}
\left[1+\frac{x(M_0^2-2m_1^2-2m_2^2)}{2Q^2(1-x)}+\cdots\right]
\nonumber\\
p_{10}&=&\sqrt{m_1^2+p_z^2}=
\frac{Q}{2}\sqrt{\frac{1-x}{x}}
\left[1+\frac{x(M_0^2+2m_1^2-2m_2^2)}{2Q^2(1-x)}+\cdots\right]
\nonumber\\
p_{20}&=&\sqrt{m_2^2+p_z^2}=
\frac{Q}{2}\sqrt{\frac{1-x}{x}}
\left[1+\frac{x(M_0^2-2m_1^2+2m_2^2)}{2Q^2(1-x)}+\cdots\right]
\, .
\end{eqnarray}
There are two possibilities corresponding to the two terms in
the current (\ref{5eq1b}).  The first term in the current will
be large only if the momentum of particle 1 is in the
direction of $q_3$ (i.e.~$p_{1z}=|p_z|>0$).  [Choosing it in
the opposite direction requires a very large momentum flow
through the wave function, and is suppressed.]  In this case  
\begin{eqnarray}
p^{(1)}_{1+}&=&Q\sqrt{\frac{1-x}{2x}}
\left[1+\frac{x(M_0^2-2m_2^2)}{2Q^2(1-x)}+\cdots\right]
\nonumber\\
p^{(1)}_{1-}&=&\sqrt{\frac{x}{2(1-x)}}\;
\frac{m_1^2}{Q}
\left[1-\frac{x(M_0^2-2m_2^2)}{2Q^2(1-x)}+\cdots\right]
\nonumber\\
p^{(1)}_{2+}&=&\sqrt{\frac{x}{2(1-x)}}\;
\frac{m_2^2}{Q} 
\left[1-\frac{x(M_0^2-2m_1^2)}{2Q^2(1-x)}+\cdots\right]
\nonumber\\
p^{(1)}_{2-}&=&Q\sqrt{\frac{1-x}{2x}}
\left[1+\frac{x(M_0^2-2m_1^2)}{2Q^2(1-x)}+\cdots\right]
\, , \label{5eq5a}
\end{eqnarray}
where the superscript ${(1)}$ is a reminder that these
relations hold only for the {\it first\/} term in the current.
The momentum relations for the {\it second\/} term in the
current (\ref{5eq1b}) follow by interchanging 1 and 2 on both
sides of these equations.  We see that the first term in the
current gives
$p_{1+}\to Q$ and
$p_{2+}\to 1/Q$, while the second term gives $p_{1+}\to 1/Q$
and $p_{2+}\to Q$.  The two terms describe kinematically
distinct regions of phase space, and their interference can
safely be neglected when the total cross section is computed.
Using the definitions (\ref{defza}) and (\ref{defzp}), $z$
and $z'$ can be easily related to the Bjorken variable $x$. 
We have
\begin{eqnarray}
z&=&1-\frac{p^{(1)}_{2-}}{P_-}\simeq 1-(1-x)=x \nonumber\\
z'&=&1-\frac{p^{(2)}_{1-}}{P_-}\simeq 1-(1-x)=x
\, .
\end{eqnarray}

\section{The bound state transition currents}
\label{appen:D}

This Appendix includes details of the evaluation of the
transition currents shown in Fig.\ \ref{hadronicamplitude}.  
The $e_1$ term will be evaluated first. The $e_2$ term can
then be obtained by a simple substitution, as discussed at the end
of this Appendix. 

The $e_1$ term, given in Eq.\ (\ref{3eq1-}), is reduced by
introducing the momentum fractions  
\begin{equation}
\xi=\frac{(P+k)_-}{P_-}
\qquad\eta=\frac{(P_{f}+k)_-}{q_-}
\qquad y=-\frac{q_-}{P_-}\, , \label{frac2}
\end{equation}
so that  
\begin{equation}
\frac{P_{f-}}{P_-}=1-y\qquad\frac{(P+k)_-}{q_-}=\eta-1
\qquad\xi'=\frac{(P_{f}+k)_-}{P_{f-}}=\frac{\xi-y}{1-y} \, .
\label{frac3}
\end{equation}
In this notation the $x$-type transverse quark current,
evaluated in Eq.~(\ref{ff6}), is 
\begin{eqnarray}
j_1^x&=&\gamma_x  -
\gamma_x\,\gamma_-\left\{\frac{g^2}{2\pi\,m_{01}}\;
\frac{q_-}{(P_{f-}+k_-)(P_-+k_-)}-\frac{q_+}{m_{01}}
F_1(\eta,Q^2)\right\}\nonumber\\
&=&\gamma_x  + \gamma_x\,\gamma_-\,H_1(\eta,q)\, ,
\end{eqnarray} 

First evaluate the
numerators, $N^\mu$, of the traces in (\ref{3eq1-}).
The numerator $N_x$ is 
\begin{eqnarray}
N_x&=&Tr\left[\gamma_-\gamma_x
\Bigl\{m_{01}+(P_f+k)_-\gamma_+\Bigr\}\left(\gamma_x 
+\gamma_x\,\gamma_-\,H_1(\eta,q)\right)\right.\nonumber\\
&&\qquad\times\left.\Bigl\{m_{01}+(P+k)_-\gamma_+\Bigr\}
\gamma_- \,(k_-\gamma_+)\right]\nonumber\\
&=&16\,k_- \,\left[m_{01}\,q_- +2(P_f+k)_-(P+k)_-
H_1(\eta,q)\right]\nonumber\\
&=&16\,k_-q_-
\frac{m_1^2}{m_{01}}-32(P_f+k)_-(P+k)_-\frac{q_+}{m_{01}}\,
F_1(\eta,Q^2)\nonumber\\
&=&\frac{16k_-q_-}{m_{01}}\left(m_1^2-
\eta(1-\eta)\,Q^2 \,F_1(\eta,Q^2)\right)\, , \label{nx}
\end{eqnarray} 
where the trace has been evaluated in a Dirac space of four
dimensions. The numerator of the minus component of the current
has only one term
\begin{eqnarray}
N_-&=&Tr\Bigl[\gamma_-
(P_f+k)_-\gamma_+\gamma_- (P+k)_-\gamma_+
\gamma_-
\,(k_-\gamma_+)\Bigr]\,\left[1+F_1(\eta,Q^2)\right]\nonumber\\
&=&32\,k_- (P_f+k)_-(P+k)_-\,\left[1+F_1(\eta,Q^2)\right]
\nonumber\\
&=&16k_-\,\frac{q_-}{q_+}\,\eta(1-\eta)\,Q^2\,
\left[1+F_1(\eta,Q^2)\right]\,  . \label{minus}
\end{eqnarray} 
The trace for the plus component is
\begin{eqnarray}
N_+&=&Tr\left[\gamma_-
\Bigl\{m_{01}+(P_f+k)_-\gamma_+\Bigr\}\left(\gamma_+ 
+\gamma_-\,G_1(\eta,q)\right)\right.\nonumber\\
&&\qquad\times\left.\Bigl\{m_{01}+(P+k)_-\gamma_+\Bigr\}
\gamma_- \,(k_-\gamma_+)\right]\nonumber\\
&=&16k_-m_1^2-32k_-(P_f+k)_-(P+k)_-\,\frac{q_+}{q_-}\,
F_1(\eta,Q^2) \nonumber\\
&=&16k_-\left(m_1^2-\eta(1-\eta)\,Q^2\, F_1(\eta,Q^2)\right)\,
.\label{plusc}
\end{eqnarray} 

Inserting the numerator (\ref{minus}) into the general result
(\ref{3eq1-}) gives the following result for the minus component
of the current   
\begin{eqnarray}
&&\!\!\!\!\!\!\!\!\left<f_-|{\cal
J}_{-}(P_f,P)|i_-\right>\Big|_{e_1\,{\rm term}}\nonumber\\
&&\!\!\!\!= 32i e_1 \int \frac{d^2k}{(2 \pi)^2} 
\frac{(P_f+k)_-(P+k)_-k_-\,
G_f\left(-P_f-k,-k\right)G_i\left(P+k,k\right)}
{[m_1^2-(P_f+k)^2-i\epsilon][m_1^2-(P+k)^2-i\epsilon]
[m_2^2-k^2-i\epsilon]}\,\left[1+F_1(\eta,Q^2)\right]  
\nonumber\\ 
&&\!\!\!\!=32i e_1\int \frac{d^2k}{(2 \pi)^2} 
\;\frac{1}{d_3} G_f\left(-P_f-k,-k\right)\,G_i\left(P+k,k\right)
\,\left[1+F_1(\eta,Q^2)\right] \, ,
\label{3eq1}
\end{eqnarray}
where the subscript - on $f_-$ reminds us that the final state
must have a pure $\gamma_-$ structure (if the ground state
does), $d_3$ is the product of the denominators of the three
quark propagators, with three poles in $k_+$ 
\begin{eqnarray}
d_3=\left[\frac{m_1^2}{\xi'\,P_{f-}}-\frac{M_f^2}{P_{f-}}
-2k_+-i\epsilon_a\right]
\left[\frac{m_1^2}{\xi\,P_-}-\frac{M_i^2}{P_-}-2k_+-i
\epsilon_b\right]
\left[\frac{m_2^2}{(\xi-1)\,P_-}-2k_+-i\epsilon_c\right]\,
,\qquad
\label{3poles}
\end{eqnarray} 
and the $\epsilon$'s change sign according to 
\begin{eqnarray}
\epsilon_a&=&\epsilon/(P_f+k)_-=\epsilon/(\xi'\,P_{f-})
=\epsilon/[(\xi-y)\,P_-]\nonumber\\
\epsilon_b&=&\epsilon/(P+k)_-=\epsilon/(\xi\,P_-) \nonumber\\
\epsilon_c&=&\epsilon/k_-=\epsilon/[(\xi-1)\,P_-] \, .
\label{epsilon}
\end{eqnarray} 
Since the vertex functions $G$ do not depend on $k_+$, we can
evaluate the $k_+$ integral.  It will be nonzero only when the
three poles of (\ref{3poles}) do not all lie in the same half
plane.  Since $q_-<0$ for electron scattering,
$P_->P_{f-}>0$, which implies that  
\begin{eqnarray}
0<y<1 \, .
\end{eqnarray} 
Hence
\begin{eqnarray}
\xi-1<\xi-y<\xi \, ,
\end{eqnarray} 
and all three poles will be in the same half of the complex
plane (giving zero for the integral) unless
\begin{eqnarray}
0<\xi<1 \, .
\end{eqnarray} 
There are two terms, depending on the sign of $\xi-y$. 
Closing the $k_+$ contour in the upper
half plane, and using $dk_-=P_-\,d\xi$ and
$(\xi-1)P_-=(\xi'-1)P_{f-}$ gives   
\begin{eqnarray}
\left<f_-|{\cal
J}_{-}(P_f,P)|i_-\right>\Big|_{\scriptsize\begin{array}{l}
e_1\cr{\rm term}\end{array}}\!\!\!\!\!\!&=& 16 e_1\,P_-^2P_{f-}
\int_0^{1}
\frac{d\xi}{(2 \pi)}
\;\frac{G_f\left(\xi',P_f\right)\,G_i\left(\xi,P\right)}
{\Delta(\xi',P_f^2)\,\Delta(\xi,P^2)}
\left[1+F_1(\eta,Q^2)\right]
\nonumber\\    
&&-16 e_1\,P_-^2P_{f-} \int_{0}^{y}\frac{d\xi}{(2\pi)}
\;\frac{G_f\left(\xi',P_f\right)\,G_i\left(\xi,P\right)
\left[1+F_1(\eta,Q^2)\right]}
{\Delta(\xi',P_f^2)\,\left(\Delta(\xi,P^2)-\displaystyle{
\frac{\Delta(\xi',P_f^2)}{(1-y)}}
\right)}
\, ,\qquad\quad
\label{3eq1a}
\end{eqnarray}
where the first term is the contribution from the pole due to
the zero in the third term in (\ref{3poles}) above, and the
second from the pole due to the zero in the first term.  Using the
definition (\ref{2eq3d}) of the wave function and combining the
two contributions in the region
$[0,y]$ we get Eq.\ (\ref{3eq1b})  with 
\begin{eqnarray}
{\cal R} &=&
\frac{\Delta(\xi',P_f^2)}{\Delta(\xi',P_f^2)-
(1-y)\,\Delta(\xi,P^2)} \nonumber\\
&=&
\frac{\xi\,\left[\xi'(1-\xi')M_f^2-(1-\xi') m_1^2 -\xi'
m_2^2\right]} {(1-\xi')\left\{\xi\,\xi'\left(M_f^2-(1-y)M_i^2
\right)-y\, m_1^2\right\}} \quad\to 1\quad{\rm if}\quad
M_f^2\to
\infty\, . \label{Rlimit}
\end{eqnarray}
This factor is needed for positive values of $\xi$ in the
interval $[0,y]$ and {\it negative\/} values of $\xi'$ in the
interval $[-y/(1-y),0]$. In this region the denominator has a zero
only if $m_1^2<0$.  In order to avoid the discussion of such
cases we limit numerical applications to cases with
$m_1^2>0$.  

Results for the
other components of the transition current, Eq.\
(\ref{currentd}), are obtained by a similar arguement using
(\ref{nx}) and (\ref{plusc}) in place of (\ref{minus}). 

Now consider the modifications required in order to evaluate
the $e_2$ term. Using the momenta defined in Fig.\
\ref{hadronicamplitude}, and the same definitions of momentum
fractions (\ref{frac2}) and (\ref{frac3}), the
$e_2$ term is obtained from the $e_1$ term simply by substituting
$m_1\leftrightarrow m_2$ and $e_1\to e_2$.  However, the momentum
fractions in the wave functions are, by convention, the fraction
of the momentum carried by the quark $m_1$, and the momentum
fraction in the quark form factor is that of the the {\it
outgoing\/} quark, and hence these functions must be written in
terms of  
\begin{eqnarray}
\xi_2&=&
\frac{-k_-}{P_-}=1-\xi\nonumber\\
\xi'_2&=&
\frac{-k_-}{P_{f-}}=1-\xi'\nonumber\\
\eta_2&=&
\frac{-(P+k)_-}{q_-} = 1-\eta\, ,
\end{eqnarray}
These observations lead immediately to the final result
(\ref{exacttran}).   


\section{The Modified Cubic Splines}
\label{appen:splines}

The two body equations were solved using a modification of the
standard cubic splines employed previously in many problems.  The standard splines are defined on 4 segments of length $h$, bounded by the 5 points $a=(n-2)h, b=a+h, c=b+h, d=c+h, e=d+h$: 
\begin{eqnarray}
S_n(x)=\frac{1}{4}\cases{\frac{(x-a)^3}{h^3}  & 
 if $a<x<b$\cr
 1+3\left[\frac{(x-b)}{h}+\frac{(x-b)^2}{h^2}-\frac{(x-b)^3}{h^3}\right] &  
 if $b<x<c$\cr
 1+3\left[\frac{(d-x)}{h}+\frac{(d-x)^2}{h^2}-\frac{(d-x)^3}{h^3}\right]  &   
 if $c<x<d$\cr
\frac{(e-x)^3}{h^3}   &  
if $d<x<e$\, .}
\end{eqnarray}
If the interval [0,1] is spanned by $n_s-2$ standard splines, there must be $n_s+1$ segments, of length $h=1/(n_s+1)$.  The first spline is numbered $n=2$ beginning at $x=0$ and the last is numbered $n_s-1$ ending at $x=1$.

\begin{figure}
\leftline{
\mbox{
   \epsfysize=2.9in\epsffile{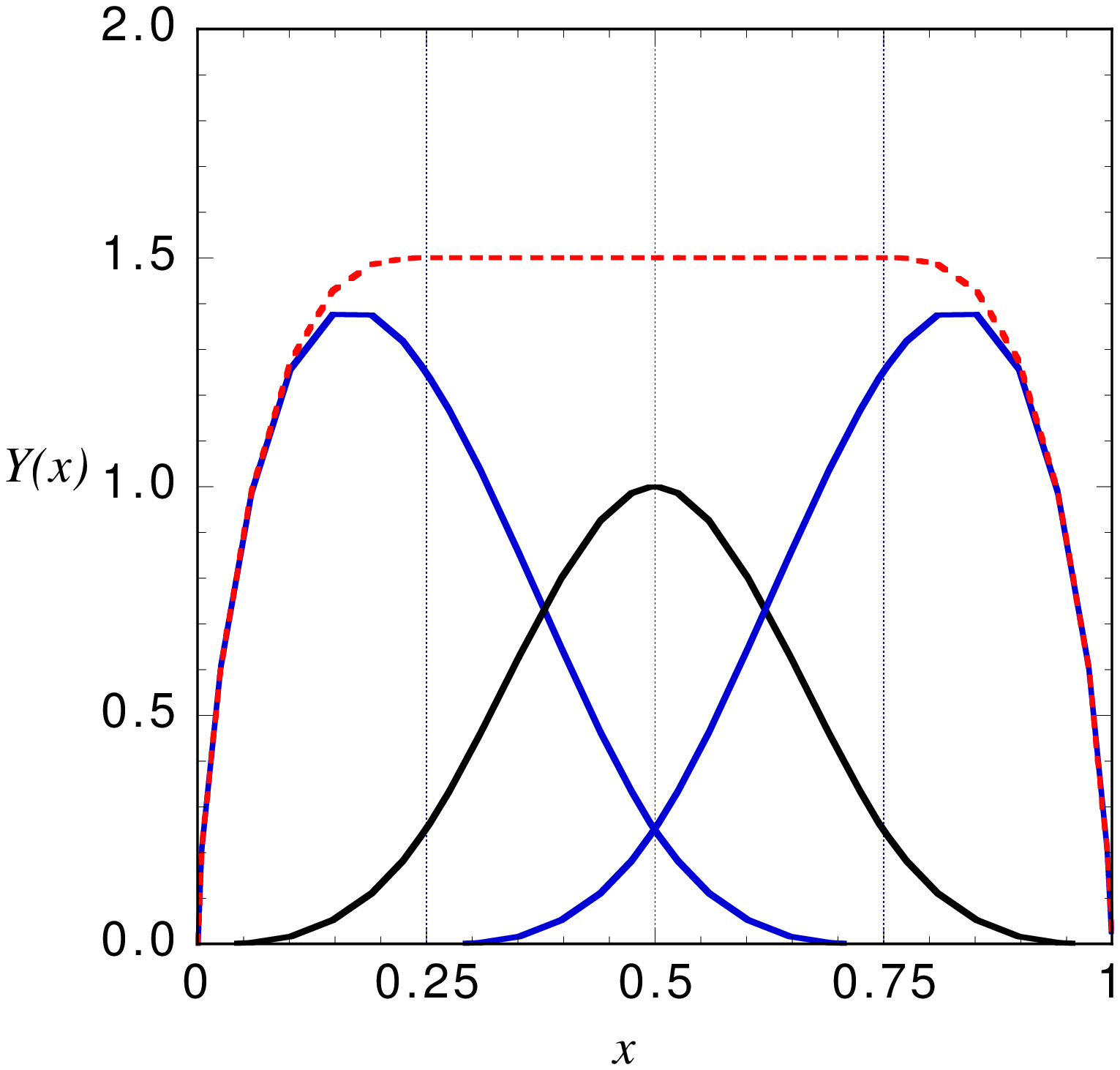}
}
}
\vspace*{-2.75in}
\rightline{
\mbox{
   \epsfysize=2.8in\epsffile{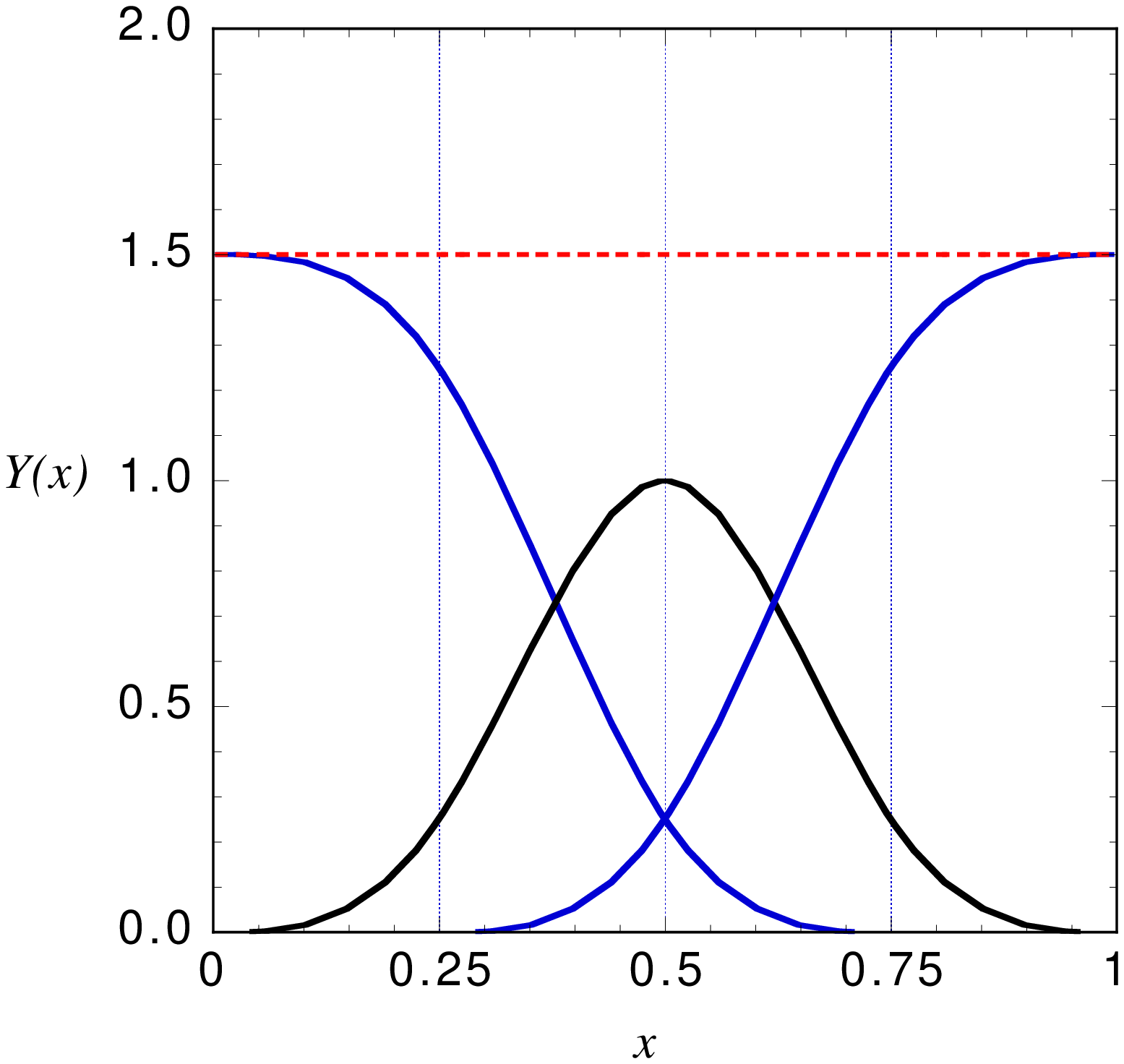}
}
}

\caption{\footnotesize\baselineskip=10ptThe modified cubic splines for the minimal case of
$n_s=3$, with four segments.  In the left figure the left- and
right-most splines at the boundary approach 0  as
$x^{0.6705}$ and 
$(1-x)^{0.6705}$ appropriate to quark masses
$m_{01}=m_{02}=1.5$.  In the right figure the bare quark masses
are zero and the splines approach a constant at the boundaries. 
Note that the sum of the splines (the dotted line) is a constant
in the central region, and throughout the {\it whole\/} region in
the chiral case. }
\label{fig:spline}
\end{figure}

The modified splines used in this paper consist of the standard splines plus one additional, non-standard spline inserted at the beginning and end of the interval [0,1].  These non-standard splines are defined on only three segments, and will be numbered 1 and $n_s$.  They are   
\begin{eqnarray}
S_1(x)&=&\cases{
 d_1\frac{x^{\beta_1}}{h^{\beta_1}}+d_2\frac{x^2}{h^2}+ d_3\frac{x^3}{h^3}  &   if $0<x<h$\cr
 \frac{1}{4}  + \frac{3}{4}\left[\frac{(2h-x)}{h}+\frac{(2h-x)^2}{h^2}
 -\frac{(2h-x)^3}{h^3}\right]  &   
 if $h<x<2h$\cr
\frac{(3h-x)^3}{4h^3}   &  
if $2h<x<3h$} \nonumber\\
S_{n_s}(x)&=&\cases{\frac{(x-1+3h)^3}{4h^3}  
 &   if $1-3h<x<1-2h$\cr
 \frac{1}{4}  + \frac{3}{4}\left[\frac{(1-h-x)}{h}+\frac{(1-h-x)^2}{h^2}
 -\frac{(1-h-x)^3}{h^3}\right]  &   
 if $1-2h<x<1-h$\cr
d_1\frac{(1-x)^{\beta_2}}{h^{\beta_2}}+d_2\frac{(1-x)^2}{h^2}+ d_3\frac{(1-x)^3}{h^3}   &   if $1-h<x<1$\, .} 
\end{eqnarray}
where the $\beta_i$ are the fractional exponents determined by the boundary conditions [given in Eq.\  (\ref{betas})], and the coefficients $d_i$ depend on the exponents $\beta$
\begin{eqnarray}
d_1=\frac{3}{(3-\beta)(2-\beta)}\, ,\qquad
d_2=\frac{3(1-\beta)}{(2-\beta)}\, ,\qquad
d_3=\frac{(2\beta-3)}{(3-\beta)}\, ,
\end{eqnarray}
where $\beta=\beta_1$ for $S_1$ or $\beta_2$ for $S_{n_s}$.  These coefficients were fixed by the requirement that the spline and its first two derivatives be continuous.  Figure \ref{fig:spline} shows two examples of the modified splines for the minimal number $n_s=3$.  In this case the central spline spans the full interval [0,1] and has 4 segments. 

Using these splines as a basis, the equation is reduced to a
matrix equation that is solved with 
the standard eigenvalue subroutine packages.

\end{document}